\documentclass[12pt]{article}

\usepackage{newtxtext,newtxmath}

\usepackage{graphicx}

\usepackage[letterpaper,margin=1in]{geometry}

\renewenvironment{abstract}
	{\quotation}
	{\endquotation}

\date{}

\makeatletter
\renewcommand{\fnum@figure}{\textbf{Figure \thefigure}}
\renewcommand{\fnum@table}{\textbf{Table \thetable}}
\makeatother

\usepackage[authoryear,round]{natbib}

\usepackage{url}

\usepackage[hidelinks]{hyperref}

\usepackage{booktabs}
\usepackage{comment}
\usepackage{float}
\usepackage{setspace}
\usepackage{soul}

\usepackage[resetlabels]{multibib}
\newcites{App}{Supplementary Material references}

\def\scititle{A Quarter of US-Trained Scientists Eventually Leave. Is the US Giving Away Its Edge?}

\title{\bfseries \boldmath \scititle}

\author{
    Dror~Shvadron$^{1\dagger}$,
    Hansen~Zhang$^{2\dagger}$,
    Lee~Fleming$^{3\dagger}$,
    Daniel~P.~Gross$^{2,4\ast\dagger}$ \and
    \small$^{1}$Rotman School of Management, University of Toronto, Toronto, Canada.\and
    \small$^{2}$Fuqua School of Business, Duke University, Durham, NC, USA.\and
    \small$^{3}$Haas School of Business, UC Berkeley, Berkeley, CA, USA.\and
    \small$^{4}$National Bureau of Economic Research, Cambridge, MA, USA.\and
    \small$^\ast$Corresponding author. Email: daniel.gross@duke.edu\and
    \small$^\dagger$First two authors listed in order of contribution. Final two authors listed alphabetically.
}

\begin{document} 

\maketitle

\begin{abstract}
\bfseries \boldmath

Using newly-assembled data from 1980 through 2024, we show that 25\% of scientifically-active, US-trained STEM PhD graduates leave the US within 15 years of graduating. Leave rates are lower in the life sciences and higher in AI and quantum science but overall have been stable for decades. Contrary to common perceptions, US technology benefits from these graduates' work even if they leave: though the US share of global patent citations to graduates' science drops from 70\% to 50\% after emigrating, it remains five times larger than the destination country share, and as large as all other countries combined. These results highlight the value that the US derives from training foreign scientists---not only when they stay, but even when they leave.

\end{abstract}

\noindent

\pagebreak

The US higher education system trains many of the world's new scientists, including foreign nationals \cite{stephan2001exceptional, freeman2010globalization}, prompting concerns that US taxpayers subsidize the training of other nations' STEM workforces \cite{barteau2024international}. Using new data which tracks US-trained STEM PhDs through 2024, we show that despite foreign nationals comprising nearly 50\% of trainees, only 10\% leave the US within five years of graduating, and only 25\% within 15 years. These rates vary across fields and subjects which are US national security priorities (e.g., quantum or AI), by destination, and over time. Stayers are more highly cited by global science and patents and more likely to make a commercially valuable discovery, as indicated by associated patents. The US share of citations made in patents to US PhD graduates' scientific publications drops from 70\% to 50\% if the scientist leaves the US, but contrary to popular assumptions remains as large as that of all other countries combined, while destination country assignees' share of citations to US PhD graduates' science increases from 2\% to 10\% when these scientists immigrate. These findings illustrate one of the ways the US benefits from training foreign STEM students, and the additional value in ensuring they stay.

By focusing on active scientists, studying the downstream consequences of emigration on US innovation, and measuring PhD graduates across four decades to identify trends, we expand on prior research studying the international mobility of US-trained STEM PhD graduates \cite{gaule2014comes, franzoni2015international, ganguli2019will, kahn2016important, kahn2020impact}. Our work is arguably closest to recent analyses of foreign national PhD graduates' intent-to-stay at the time of graduation \cite{zwetsloot2020trends} and subsequently-realized stay rates \cite{corrigan2022long} in National Science Foundation (NSF) surveys of US PhD graduates. Yet even these analyses are limited by the underlying surveys' timing, sampling, and response rates and cannot be linked to respondents' science or its impacts.\footnote{Studies of intent-to-stay are based on responses to the NSF's Survey of Earned Doctorates (SED), which are recorded before outcomes are realized. Studies of long-term stay rates are typically based on NSF's biennial Survey of Doctorate Recipients (SDR), which samples around 10\% of past graduates in each survey wave and has lower sampling and higher non-response rates from graduates residing outside of the US, limiting the quality of the inference that can be drawn (see Corrigan et al. 2022 for discussion). Additionally, because these studies typically examine only one survey wave, they only observe each cohort at a single moment in time and are thus unable to make longitudinal comparisons of short- and long-term stay rates, as we will in Figure 1.} Other research which has studied US PhD graduates' migration patterns typically evaluates smaller and more focused samples, due to data constraints at the time. Using new AI-based linking methods, we build on this literature with an updated, near-population sample of graduates linked to their post-PhD research and to technology that builds on it, which enables us to evaluate graduate mobility and its consequences at very fine strata and over long horizons.

\subsection*{Methods and Data}

To study migration patterns, we created a new dataset linking US-trained PhD graduates, scientific publications, and patents.\footnote{See SM for details of data sources and methods, including on the implementation and validation of the crosswalks and machine learning models used in classification and record linking.} Our core sample comprises STEM PhD graduates from US universities in ProQuest Dissertations \& Theses Global (PQDT) between 1980 to 2019. This sample closely aligns with totals in the National Science Foundation's (NSF) Survey of Earned Doctorates (SED) (an annual census of US PhD graduates) both over time and by institution and subject, indicating that it approximates the SED population.\footnote{There is some modest attrition from PQDT in recent years. See SM for discussion.} Using titles and abstracts from graduates' dissertations and a custom, validated LLM-based classifier, we additionally categorize graduates to 18 ``critical and emerging technology areas'' identified by the White House Office of Science and Technology Policy (OSTP) as US national and economic security priorities.

We trace graduates' scientific careers by identifying their subsequent publications. Longitudinal linking is challenging because of the absence of shared identifiers across data sources and the prevalence of non-unique names---a difficulty that has in the past hampered population-level analyses of PhD graduates' careers and constrained prior research to studying smaller subpopulations that could be manually linked. Advances in computational tools such as high-dimensional textual embeddings, LLMs, and cloud computing have recently made large-scale linkage feasible. We develop a custom two-stage supervised machine-learning model trained on self-reported career histories in ORCID, a public registry where scholars can identify themselves, their training, and their publications. Our approach achieves high precision and recall across ethnicities while maintaining transparency in performance (see SM for complete methodological details).

Linking PhD graduates to their future science enables us to identify institutional affiliations over time, and in turn geographic location. By observing the country where a scientist is employed, we infer emigration and its timing. Recently-developed data by Marx and coauthors additionally enable us to measure the incorporation of science in technology \cite{marx2020reliance}. We thus evaluate graduates' contributions vis-\`{a}-vis (i) their publications, (ii) citations to those publications by other scientific articles, (iii) citations by patents, and (iv) patents issued on their research (``patent-paper pairs''), which indicate that a discovery had immediate commercial value.

\subsection*{International emigration rates and destinations of US trained STEM PhDs}

Panel A of Figure 1 documents the share of actively-publishing STEM PhD graduates from US institutions since 1980 that we observe publishing in other countries 5, 10, and 15 years after graduation (red, green, and yellow lines, respectively). As a reference point, we also show the share of PhD graduates in each cohort who are foreign nationals (dotted black line), as reported in the SED. Postgraduate emigration rates after 5, 10, and 15 years are approximately 10\%, 20\%, and 25\%. These emigration rates are consistent with survey evidence on graduates' intentions to stay and reported emigration from the SED and NSF Survey of Doctorate Recipients (SDR) \cite{zwetsloot2020trends, corrigan2022long} as well as an earlier hand-collected sample of chemistry PhDs \cite{gaule2014comes}. Emigration rates across PhD cohorts correlate strongly with the foreign national share of graduates ($\rho=0.86$ to $0.95$), suggesting (intuitively) that many emigrating graduates are foreign nationals, though the emigrating population in our data also includes US citizens who move abroad. In all cohorts, the 5-year (15-year) emigration rate is approximately 25\% (50\%) of the foreign national share, which has grown from around 25\% of graduates in 1980 to between 40\% and 45\% since the early 1990s.

Panel B of Figure 1 documents where these emigrants go, and how destinations have changed over time. We focus on short-run (5-year) emigration patterns, which can be observed through the 2019 cohort. Nearly all of the increase in the 5-year emigration rate since the 1980s is due to an increase in graduates moving to Asian countries. Also notable is a near-tripling of emigration to Middle Eastern countries and a halving of emigration to non-US North American countries (mainly Canada). South Korea and Taiwan were the most common destinations in Asia in the 1990s---largely fueled by semiconductor scientists. China has been the most common destination since the mid-2000s. For 2015-19 graduates (the most recent cohorts for which 5-year emigration rates are available), approximately half of emigrating graduates go to Asia, one third to Europe, and the rest to other locations. Among those who move to East and South Asian countries, 40\% move to China, 9\% to South Korea, 6\% to Japan, and 3\% to India.

Panel C of Figure 1 tabulates emigration rates by broad field, as determined from the (self-reported) subject of each graduate's dissertation in PQDT using using SED definitions and crosswalks. Roughly 30\% of graduates between 2000 and 2009 in mathematics, computer sciences, and engineering left the US within 15 years of graduating. In additional analysis at the more granular level of SED major fields, we find that over 40\% of civil and industrial engineers leave the US within 15 years, whereas less than 20\% of biology, biomedical engineering, and health science PhDs do so. Most other engineering disciplines (e.g., electrical, mechanical, chemical, and materials science) are in the 30-40\% range. The most common destination region for emigrating engineering PhD graduates is Asia, whereas the most common region for emigrating graduates in mathematics, physics and astronomy, earth sciences, and biomedicine is Europe.

In the lower half of Panel C we evaluate emigration rates among graduates whose dissertation science is related to specific critical and emerging technology areas which represent US national and economic security priorities.\footnote{Results for all 18 OSTP critical and emerging technology areas are provided in the SM.} Almost a third of quantum science and technology-related PhDs leave the US within 15 years, and between 20-30\% do so in materials science, semiconductors and microelectronics, artificial intelligence, and space technology. At the other end, less than 20\% of biotechnology-related graduates leave the US within 15 years, possibly reflecting the composition of the trainee pool (less international) or the size of the US market.

In additional analysis we also compare emigrating and non-emigrating scientists on their publications, citations from scientific articles and patents, and patent-paper pairs using a regression framework (see SM for descriptive statistics and estimated models). Graduates who stay in the US at least five years after graduation were on average more productive and impactful in their first five years of their career: stayers on average have 2.5\% more publications and 3.7\%, 10.4\%, and 16.7\% more scientific citations, patent citations, and patent-paper pairs per publication, conditional on any, and are more likely to have them at all ($p<0.001$ in all cases). The evidence suggests stayers are positively selected, though we cannot fully rule out the possibility that being in the US is instead what makes scientists more productive early on in their career \cite{kahn2016important}.

\begin{figure}[htbp]
\centering
\includegraphics[width=0.8\linewidth]{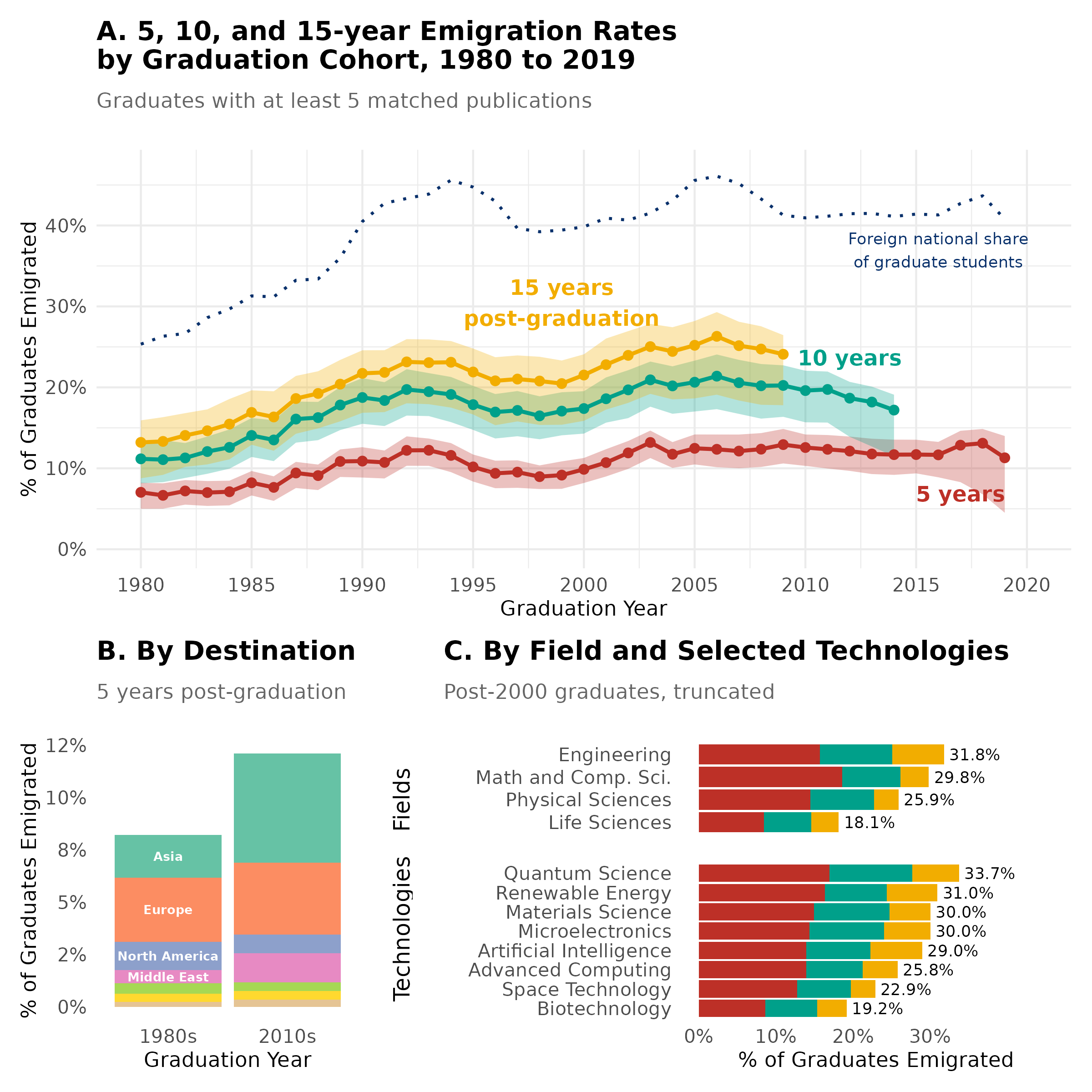} \\
\caption{\textbf{\small Approximately 12\% of US-trained STEM PhDs leave the US within 5 years of graduating and 25\% within 15 years. Leave rates track the foreign national share over time but vary by field and technology. A growing share of graduates emigrates to Asia.} Panel A shows the foreign national share of STEM PhD graduates since 1980 (black dotted line, based on the Survey of Earned Doctorates) and share of all graduates we identify as publishing in a foreign country 5, 10, and 15 years post-PhD (red, orange, and yellow lines, respectively, based on our PQDT graduate sample linked to future publication affiliations). Colored bounds represent confidence intervals by varying model specifications. Panel B shows the share emigrating to the specified global regions after 5 years in the 1980s vs. 2010s. Panel C shows the share emigrating by broad field and in select technology areas. All charts condition on graduates with at least 5 linked publications.}
\label{fig:figure1}
\end{figure}

\subsection*{The US advantage in building on the science of its trainees}

We additionally investigate how emigration impacts the commercial use of US-trained graduates' science in the US and in destination countries. To do so, we measure the proportion of worldwide patent citations to graduates' scientific publications made by patent owners (a.k.a. assignees, mostly firms, but including universities and individuals) in each country. Patent citations to science can indicate both content linkages and knowledge flows \cite{bryan2020text}, and we view countries' shares of global patent citations as an indicator of which countries benefit most from US-trained graduates' science through its application, with higher values indicating greater benefit.

In Figure 2 Panel A we show the share of citations made by US assignees to emigrating graduates' science which published before vs. after leaving the US (in dashed red and solid blue lines, respectively). We chart both shares over time, by patent filing year. For comparison, we also show the US share of citations over time to the science of graduates who never leave the US (dashed blue line) and to all foreign based science (dashed red line). US assignees account for approximately 70\% of global patent citations to all pre-emigration science by emigrating PhD graduates, closely tracking the share made to other US science. This share falls to 50\% for citations to post-emigration science---similar to the US share of citations to all foreign science. We interpret the 20\% difference in shares as an indication of the loss of absorptive capacity that takes place when a US-trained PhD graduate leaves the US; it offers a rough estimate of the importance of a scientist's physical presence in the US to the uptake of their research by US firms. Though this decline is large, that the US continues to account for half of global citations to graduates' science (and other foreign science) indicates that despite migration, much of this science \emph{nevertheless benefits the US} due to the commercialization capabilities of the US innovation system.

Panel B of Figure 2 shows the complement: the annual share of patent citations to emigrating graduates' science from assignees in the destination country. We plot these shares by region (North America, Europe, Asia, and Middle East; color-coded) and overall, with dashed lines representing destination countries' share of citations to graduates' pre-immigration science and solid lines representing their share of citations to post-immigration science. Though destination countries' patents rarely cite US-trained graduates' pre-immigration science, their share of all global citations to graduates' post-immigration science increases to roughly 10\%.

\begin{figure}[htbp]
\centering
\includegraphics[width=\linewidth]{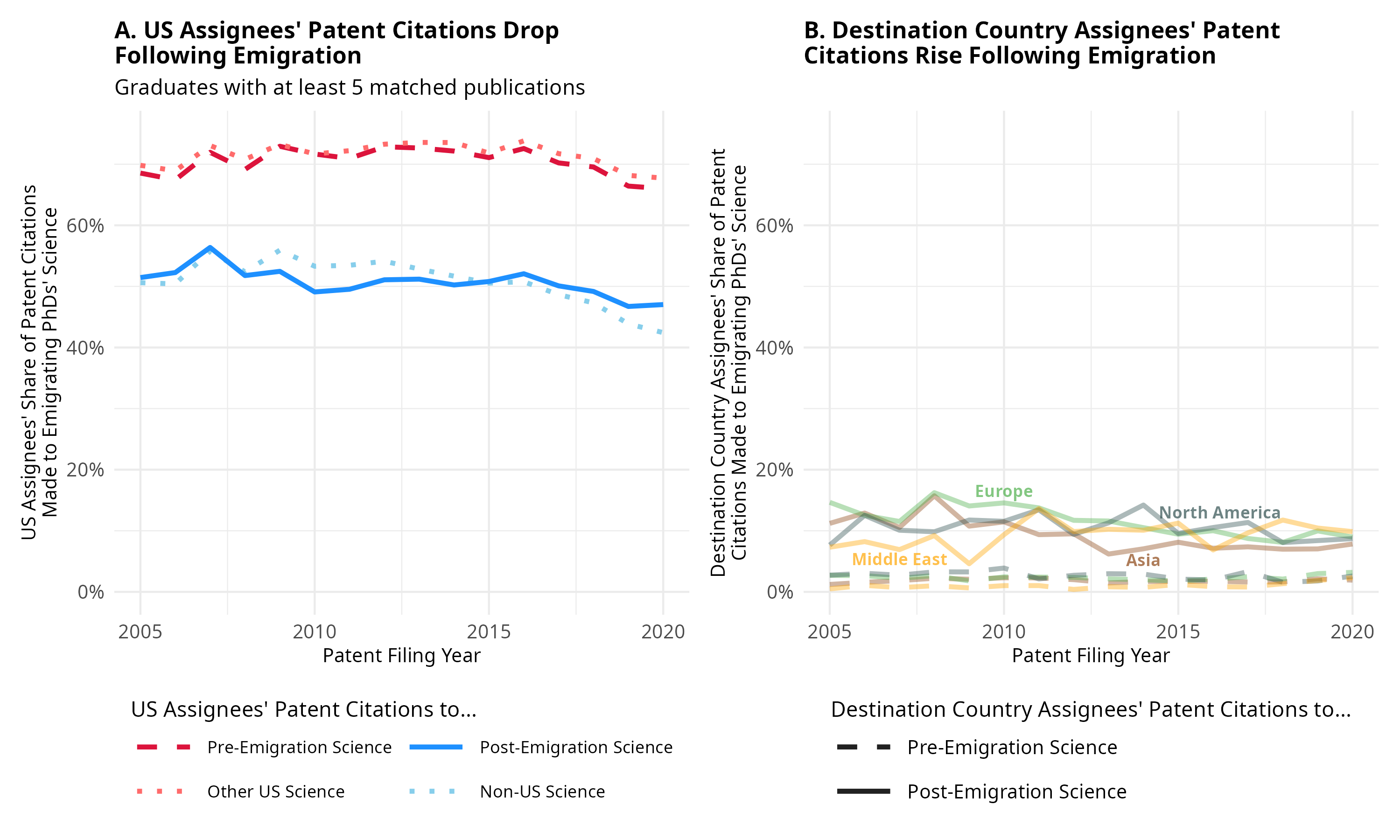}
\caption{\textbf{Post-emigration science is less likely to be cited in US commercial innovation and more likely to be cited in destination country innovation than pre-emigration science.}
Panel A evaluates US assignees' share of citations made each year in global patents to PhD graduates' science published before vs. after they leave the US (dashed red and solid blue lines, respectively). For comparison, the figure also provides US assignees' share of citations to non-leaver graduates' science and non-US science (dashed blue and red lines). Patent assignees are organizations which file and own patents (e.g., firms, universities, individuals). The figure illustrates localization in the application of PhD graduate science: US assignees' share of citations to emigrants' pre-emigration science matches that to other domestic science ($\approx$70\%), and their share to post-emigration science matches that to other foreign science ($\approx$50\%). Panel B shows a countervailing effect in destination countries, where domestic assignees' share of citations to emigrants' pre-emigration versus post-emigration science increases from roughly 2\% to 10\%.
}
\label{fig:figure2}
\end{figure}

\subsection*{Discussion}

A recurring concern among US scientific leaders and policymakers over the past eighty years has been a shortage of domestic STEM scientists relative to national needs \cite{bush1945science, atkinson1990supply}. Foreign scientists have long filled this gap. As Nice (p. 179) observes, ``recent years have seen a growing recognition of the deep reliance of the US national security innovation base on foreign national advanced degree holders,'' who help to meet ``the science and engineering workforce needs of the US defense sector and its associated innovation base'' \cite{nice2025meeting}. Similar observations have been made of the academic and commercial sectors \cite{nsb2024stem}. Our evidence is consistent with this view: the foreign share of US STEM PhD graduates has been stable for decades at a bit over 40\%, and a large majority of graduates who remain active in science---including these foreign nationals---continue working in the US for many years, where they support US competitiveness in science and technology.

Given these findings, a corollary question is what attracts foreign graduate students to the US and leads them to stay. Prior research points to immigration policy---a subject of perennial public interest---having a large effect on stay rates \cite{kahn2020impact, glennon2024skilled}. Proponents of a more accommodating immigration policy point to a history of US science and technology benefiting from foreign talent, the disproportionate impact foreign scientists have on the innovation economy, and evidence that if firms cannot meet their needs at home, they will move operations to where this human capital can be found. Critics, on the other hand, argue that foreign trainees may crowd out domestic students, are conduits for domestic science and technology to leak to geopolitical adversaries, or mainly benefit other countries through migration. Our evidence speaks to the latter two concerns, documenting the relatively high rates at which trainees have historically stayed in the US in both the short- and long-run, and suggesting that the US makes greater use of trainees' science in new technology than foreign countries, not only when they stay but also when they leave.

In addition to concerns over a shortage of STEM scientists, US policymakers have long shouldered concerns about a long-term decline in domestic high-tech manufacturing, and more recently about China's rapid growth in high-quality scientific research---the combination of which is increasingly challenging US competitiveness in the global high-tech value chain. Despite these concerns, our evidence suggests the US maintains an advantage in the critical, intermediate juncture of translating science into technology. It also suggests that retaining this edge requires continued investment in research and training---including of foreign-born students---to sustain a capacity to produce new science and apply it towards economic, medical, and national security ends.

\pagebreak


\clearpage 

\bibliography{bibliography}
\bibliographystyle{bib_econ}

\pagebreak


\section*{Acknowledgments}

We thank Britta Glennon and audiences at the NBER Early Career Scientists meeting, Max Planck Institute for Innovation and Competition, and UC Berkeley Technology Competitiveness and Industrial Policy Center workshop for feedback on this work.

\paragraph*{Funding:}
We thank the Alfred P. Sloan Foundation (grant G-2023-21064) and the UC Berkeley Technology Competitiveness and Industrial Policy Center for financial support. Gross acknowledges support from the National Science Foundation under Grant No. 2420824.

\paragraph*{Author contributions:}\mbox{}\\
Conceptualization: D.S., H.Z., L.F., D.P.G. \\
Methodology: D.S., H.Z., L.F., D.P.G. \\
Data Curation: D.S., H.Z. \\
Analysis: D.S. \\
Writing: D.S., H.Z., L.F., D.P.G. \\
Revisions: D.S., H.Z., L.F., D.P.G. \\
Project administration: D.S., H.Z., L.F., D.P.G. \\
Funding acquisition: L.F., D.P.G.

\paragraph*{Competing interests:}
There are no competing interests to declare.

\paragraph*{Data availability:}
Data and replication package will be posted to a public repository before publication. A link to the repository will be placed here.


\subsection*{Supplementary materials}
Data and Methods \\
Supplementary Results


\pagebreak


\renewcommand{\thefigure}{S\arabic{figure}}
\renewcommand{\thetable}{S\arabic{table}}
\renewcommand{\theequation}{S\arabic{equation}}
\renewcommand{\thepage}{S\arabic{page}}
\setcounter{figure}{0}
\setcounter{table}{0}
\setcounter{equation}{0}
\setcounter{page}{1} 


\begin{center}
\section*{Supplementary Materials for\\ \scititle}

Dror~Shvadron$^{1}$,
Hansen~Zhang$^{2}$,
Lee~Fleming$^{3}$,
Daniel~P.~Gross$^{2,4\ast}$ \\
\small$^\ast$Corresponding author. Email: daniel.gross@duke.edu\\
\end{center}

\subsubsection*{This PDF file includes:}
Data and Methods \\
Supplementary Results

\pagebreak


\subsection*{Data and Methods}

In this section we describe our data sources, sampling, and methods for linking US STEM PhD graduates with their post-PhD careers. Data construction essentially involves two steps: building a core sample of graduates (from ProQuest Dissertations \& Theses Global), and connecting them to postgraduate scientific activity, including location (using data from OpenAlex and a newly-developed and validated linking procedure). Our analysis also uses data compiled from the National Science Foundation (NSF) Survey of Earned Doctorates (SED), which reports responses from annual surveys of PhD graduates in STEM fields, including country of origin.

Note that portions of this appendix---particularly those documenting our sample and our classification of PhD graduates to critical technology areas---share content in common with the data appendix of \citeApp{ShvadronZhangFlemingEtAl2025FundingUSScientific}, where we use a similar sample from the same data source to study funding sources for US scientific trainees.

\subsubsection*{Core Data Sample: ProQuest Dissertations and Theses Global}

\paragraph{Sample construction} Our core analytical sample for this paper consists of US STEM PhD graduates between 1980 and 2019, which we compile from ProQuest Dissertations \& Theses Global (PQDT). ProQuest has been acquiring and re-publishing dissertations of US PhD graduates for nearly a century. Its collection has been digitally catalogued and includes extensive metadata on each dissertation (e.g., author, institution, degree name, title, abstract, subject). PQDT data have been used in a range of recent research in the science of science, economics of innovation, and other fields studying scientific trainees \citeApp{toole2010commercializing, bikard2015exploring, buffington2016stem, jiang2023tale, antman2023innovation, arora2023effect, ShvadronZhangFlemingEtAl2025FundingUSScientific}. As we show below, its contents approximate the universe of US PhD graduates in STEM, which is a result of extensive, long-running licensing agreements it has held with many (though not all) universities across the country.

At the time of acquisition (in 2024), the PQDT database contained nearly six million global master's theses and doctoral dissertations, including in the natural sciences (i.e., physical and life sciences), engineering, humanities, social sciences, and more. For this paper, we sought to build a sample of doctoral graduates at US institutions in the natural sciences and engineering---which we term STEM fields as shorthand---between 1980 to 2019. We end our sample in 2019 because it is the last cohort for which we had at least five years of post-graduation data (our minimum window for measuring graduates' emigration) at the time of data collection.

To construct this sample, we first removed (i) non-US dissertations and (ii) non-doctoral theses and dissertations. We also manually reviewed degree names to identify those corresponding to research degrees in the natural sciences and engineering. We then crosswalk these graduates' self-reported subjects (as provided in the PQDT data) to ``major fields'' defined by the US National Center for Science and Engineering Statistics (NCSES) and restrict our sample to graduates in 17 major fields which are typically considered STEM fields.\footnote{\linespread{1} \nonfrenchspacing In early years, PQDT reports a single subject for each graduate. In later years, when multiple subjects are provided, we treat the first-listed subject as the graduate's primary subject. We map PQDT subjects to major fields primarily using crosswalks from NCSES, which it uses to map subjects to fields its own surveys.} Our final PQDT sample contains roughly 800,000  US STEM PhD graduates from 1980 to 2019.

\begin{figure}[htbp]
\centering
\includegraphics[height=2.25in]{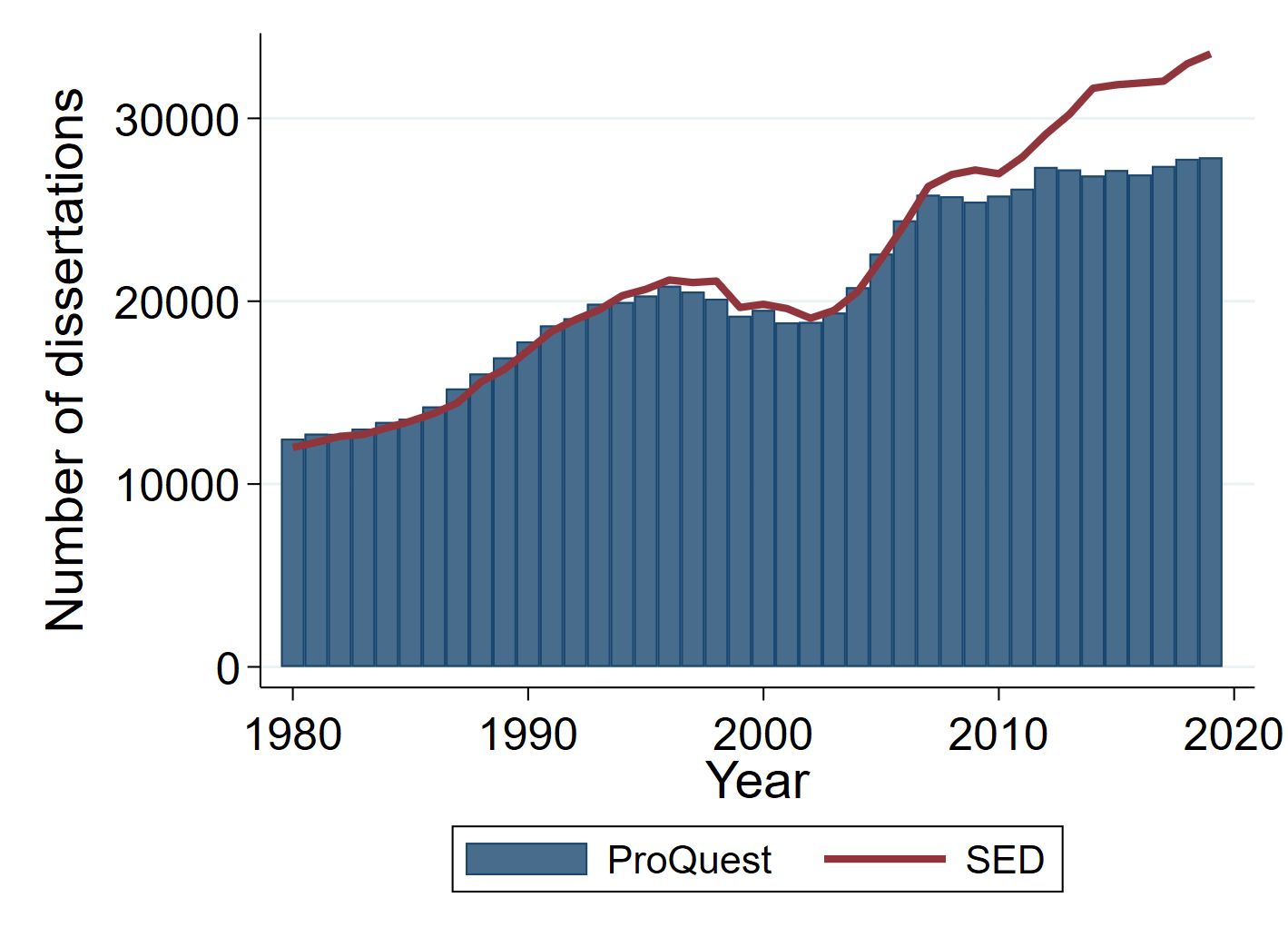}
\caption{\textbf{ProQuest vs. SED graduates in STEM fields, by year.} Figure shows annual PQDT dissertation counts in the natural sciences and engineering (blue bars) and SED counts (red line) for comparison, from 1980 to 2019. Sample consists of PQDT graduates mapped to natural science and engineering SED major fields based on their first-listed subjects.}
\label{fig:pqdt_vs_sed_totals}
\end{figure}

\begin{figure}[htbp]
\centering
\includegraphics[height=4in]{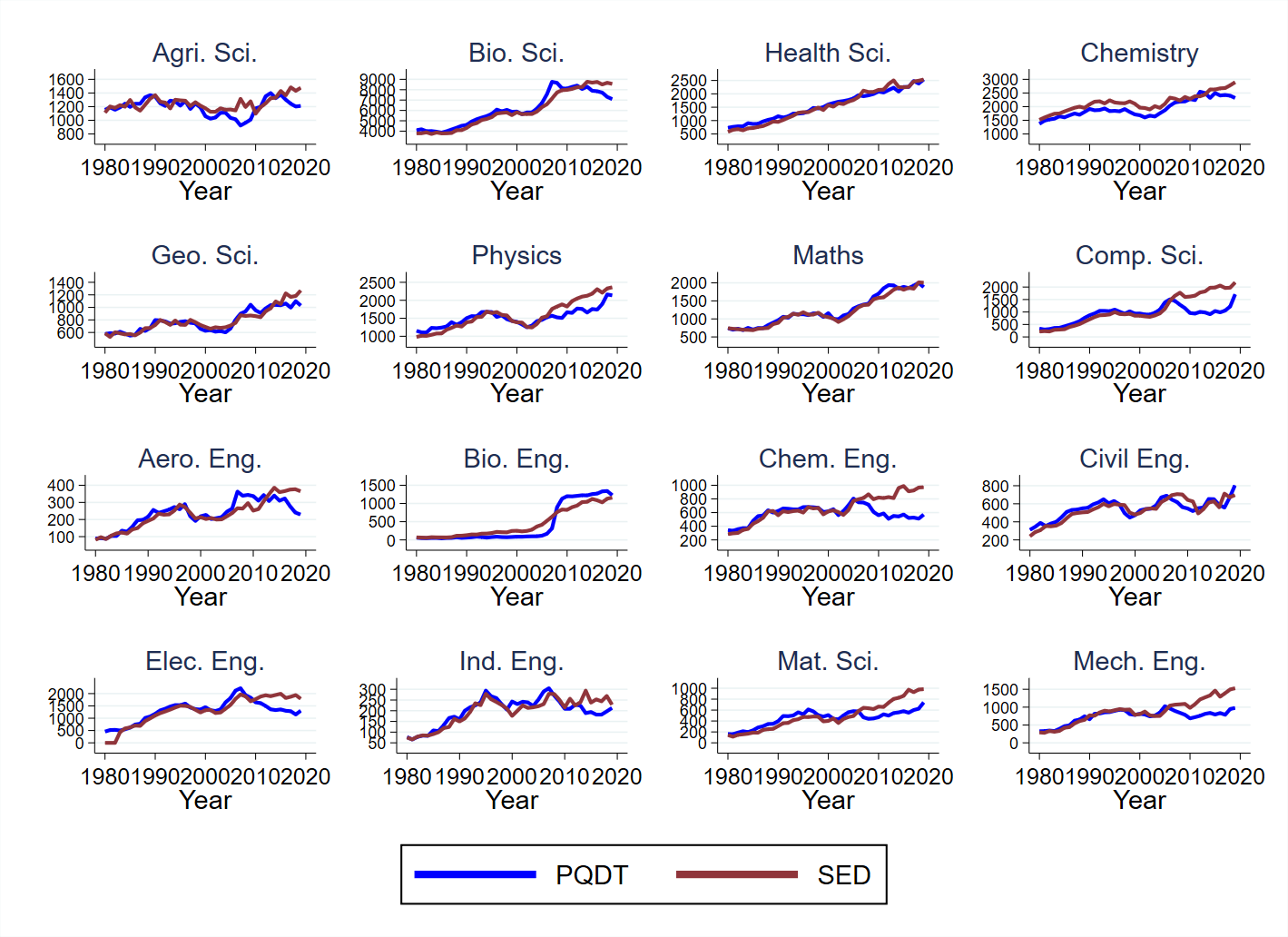}
\caption{\textbf{ProQuest vs. SED graduates in STEM fields, by subject and year.} Figure shows annual PQDT graduates in the natural sciences and engineering (blue bars) and SED counts (red line) for comparison, from 1980 to 2019, disaggregated across SED major fields. PQDT graduates mapped to SED major fields based on their first-listed subjects.}
\label{fig:pqdt_vs_sed___all_subjects}
\end{figure}

To evaluate the completeness and representativeness of our data, we compare the annual number of PQDT graduates in this sample to annual counts of US doctoral graduates in the same fields from the Survey of Earned Doctorates (SED), an annual census of research doctorate recipients from US universities which has been administered by NSF since 1958. Figure \ref{fig:pqdt_vs_sed_totals} shows that PQDT (the blue bars) tracks the SED (red line) closely to the late 2000s. Even when they diverge slightly post-2005, PQDT still sums to \textgreater90\% of the SED totals. Figure \ref{fig:pqdt_vs_sed___all_subjects} disaggregates this comparison to individual fields and shows a similar degree of consistency within them. 

\begin{figure}[htbp]
\centering
\begin{center}
\begin{tabular}{cc}
\includegraphics[height=2in]{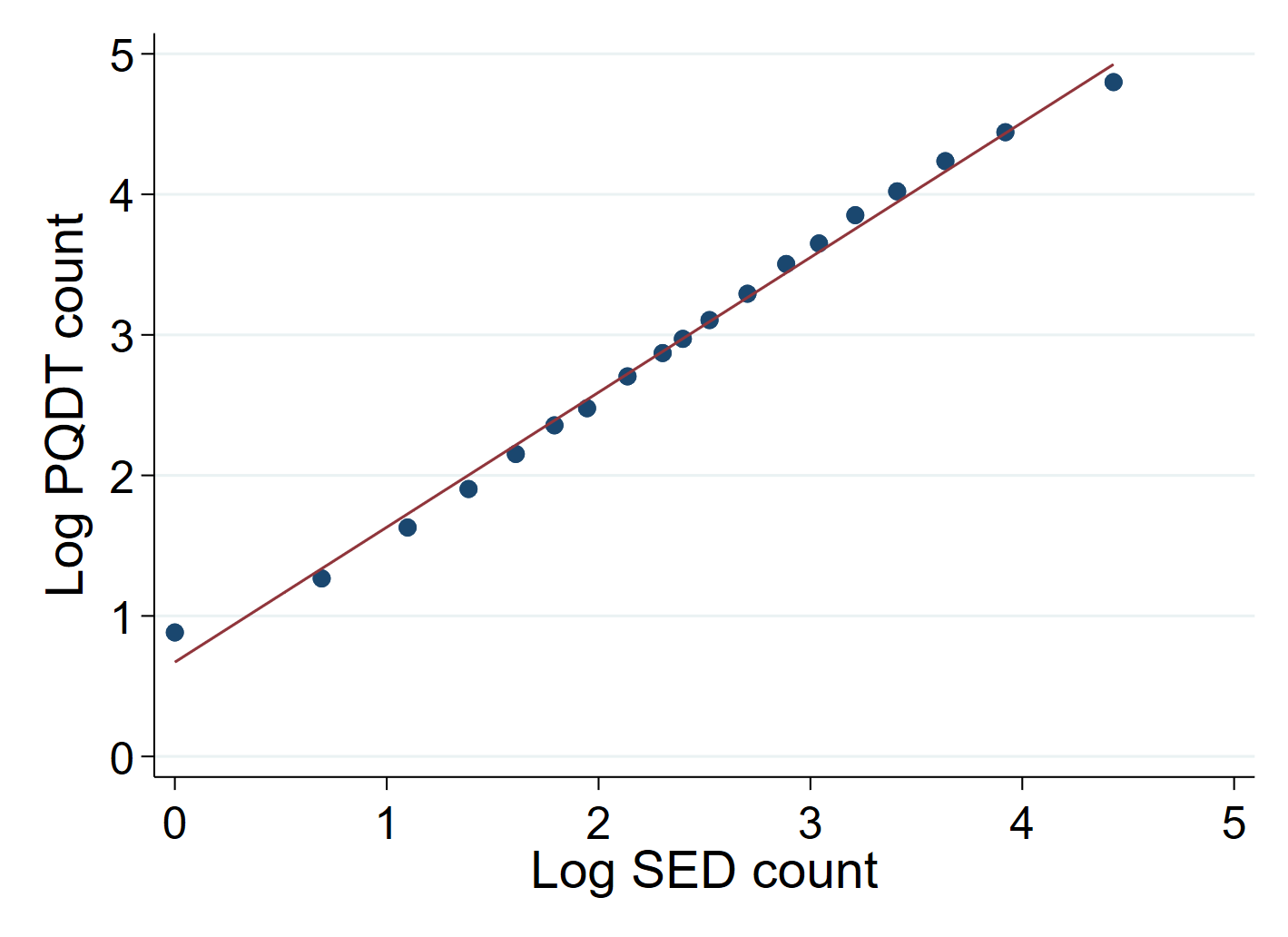} &
\includegraphics[height=2in]{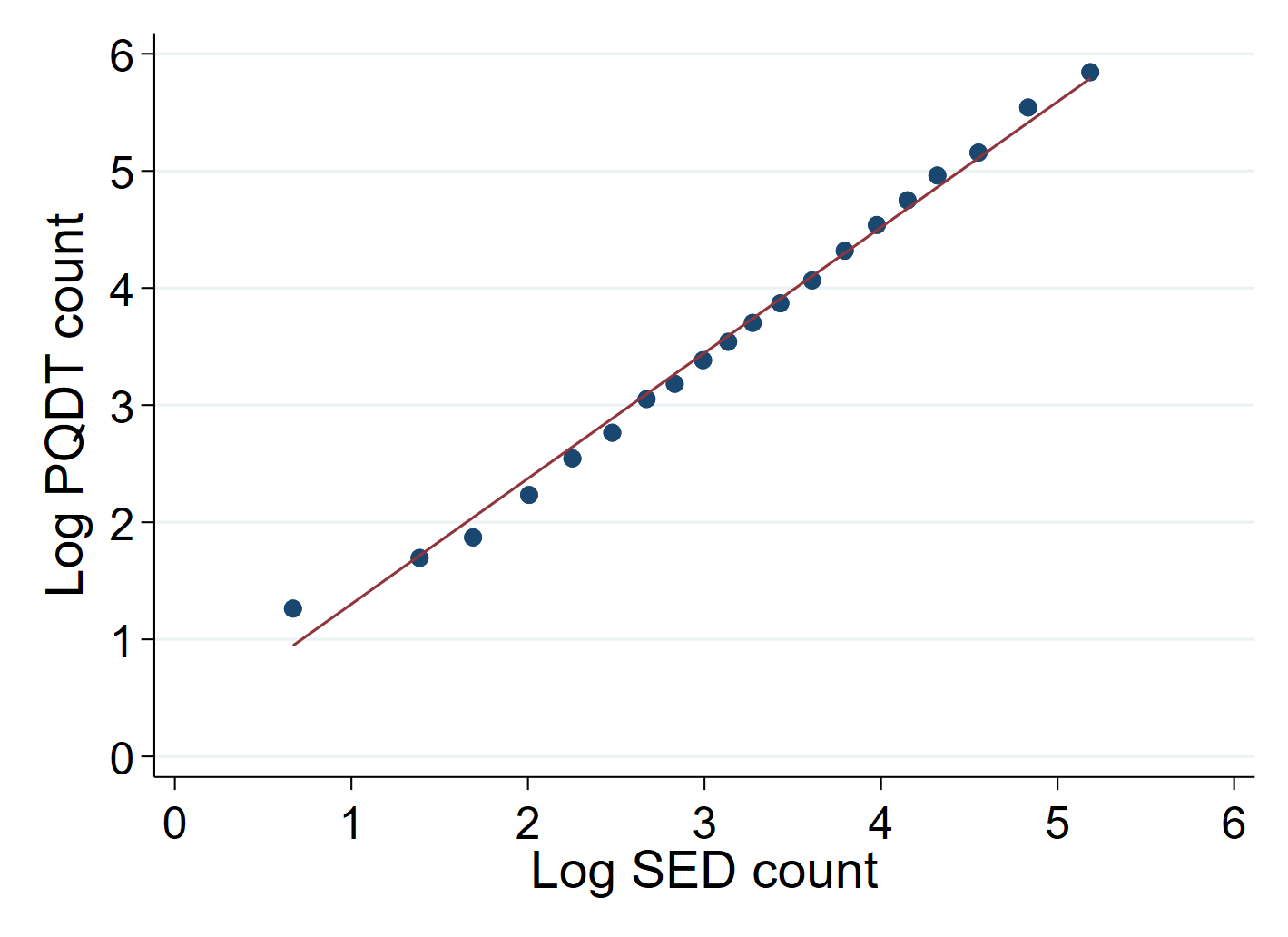} \\
$\quad$ Panel (A): University-field-year level &
$\quad$ Panel (B): University-year level
\end{tabular}
\end{center}
\caption{\textbf{Correlation of ProQuest vs. SED graduate counts at the university-field-year and university-year level.} Figure shows a binned scatterplot of log PQDT vs. log SED graduate counts at the university-year level (left panel) and university-field-year level (right panel).}
\label{fig:log_pqdt_sed__schoolfieldyear}
\end{figure}

Figure \ref{fig:log_pqdt_sed__schoolfieldyear}, Panel (A) extends these comparisons, presenting a binned scatterplot of the log number of graduates in our PQDT sample at the university-field-year level against the log number of graduates according to the SED data. Panel (B) repeats this comparison at the university-year level. The sample in these charts is by construction limited to observations with a nonzero number of graduates in both PQDT and SED---though for the most part, when one source reports zero graduates, and the other nonzero, the latter reports only one or two.

\paragraph{Sample limitations}

Collectively, this evidence suggests that PQDT approximates the universe of US-trained STEM PhDs for most of the period we study---in the neighborhood of a complete accounting pre-2005, and a 90\% accounting post-2015. Even this (modest) undermeasurement in recent years is not necessarily problematic for analysis of emigration rates, particularly if sampling is unrelated to emigration---in which case we can view results borne of a 90\% sample of graduates representative of all graduates, even as it is already close to the universe. It remains an important question, however, as to why the PQDT and SED total graduate counts have recently diverged and what this divergence implies about our sampling and inferences.

Closer examination of the PQDT sample suggests the gap is the result of a small but growing number of universities ending their relationship with PQDT, such that their graduates' data are no longer automatically included. This change appears to reflect a movement of universities towards making research open access, partly in response to federal guidance \citeApp{ostp2013memo}, and typically on university repositories in place of commercial publishers. The resulting attrition is unrelated to specific emigration patterns and thus unlikely to affect our findings.

More complete sampling is nonetheless desirable, particularly for extensions of this analysis which may examine specific institutions or programs. In ongoing work we are exploring ways to address this gap, including by collecting data from individual university repositories (one at a time), which if successful can be used both to fill residual gaps in this article sample and to extend this analysis in the future, particularly if attrition continues to grow.

\subsubsection*{Classifying PhD Graduates into Critical Technologies}

We supplement the PQDT-reported measures with measures of each graduate's association with 18 critical technology areas as specified by the White House Office of Science and Technology Policy (OSTP) in 2024 \citeApp{ostp2024critical}. The White House identified these 18 technologies as national security priorities in order to inform science and technology policy across the executive branch. This linkage serves two purposes: (i) it produces a measure of specific and immediate policy interest, enabling us to track the development of human capital across specific technology domains that are currently considered important for national security and technological competitiveness, and (ii) it also demonstrates the possibilities presented by our PQDT-based sample and textual analysis methods.

\noindent\textbf{Step 1: Zero-shot classification to technology areas}

We begin our classification with the full population of US STEM PhD graduates between 2000 and 2019. To link graduates to technology areas, we have at our disposal a range of inputs, including dissertation titles, abstracts, subjects, keywords, and even full text (see \citeApp{ShvadronZhangFlemingEtAl2025FundingUSScientific}). Rather than directly classifying PhD graduates based on these inputs directly, we first generate a concise, one-sentence summary for each dissertation using their titles and abstracts. This approach parallels recent work by Aiken et al. (2024) \citeApp{aiken2024emerging}, who employed large language models to annotate scientific publications for topic relevance in emerging technologies. Specifically, we use the \texttt{gpt-4o-mini} model with a tailored prompt designed to extract the motivation, task, and methodology of each dissertation, while excluding evaluative or impact-based language.\footnote{We use the July 18, 2024 version of the \texttt{gpt-4o-mini} model. gpt-4o-mini is optimized for high-throughput, low-latency tasks and supports both text and image inputs, though only text output was used in our application. Its affordability and responsiveness made it particularly well-suited for summarization of hundreds of thousands of dissertations. See \url{https://platform.openai.com/docs/models/gpt-4o-mini} for technical details.}

This generative summary step allows us to standardize input across a large, heterogeneous corpus of dissertations and provides a foundation for downstream classification to critical technology areas with several methodological advantages. First, it helps normalize heterogeneous textual formats: dissertation abstracts vary widely in length, structure, and verbosity, often containing extraneous information unrelated to core research contributions. Second, summarization encourages the model to foreground research motivation, task, and method, thereby aligning the input more closely with the latent conceptual criteria used in topic classification. Third, as emphasized in Aiken et al. \citeApp{aiken2024emerging}, generating structured summaries via prompt-based language models enhances both consistency and zero-shot classification performance by filtering out irrelevant content and emphasizing comparable semantic structure across documents. This preprocessing step can thus reduce noise, mitigate bias from field-specific jargon, and improve classification accuracy.

Building on these summaries, we then evaluate whether each PhD graduate's dissertation science is related to each of the 18 critical and emerging technology areas, allowing a dissertation to associate to multiple areas. For this task, we prompt \texttt{gpt-4o-mini} to act as an expert in defense and emerging technologies and make an assessment. For each PhD graduate, we supply the model with their dissertation summary, title, subject terms, and keywords, and instruct it to return a binary ``Yes/No'' judgment for each technology area, with instructions to prioritize precision over recall. The user prompt lists all technology categories and presents the metadata for one dissertation at a time. %
This approach enables us to assess dissertations in a scalable and consistent way. The end result is a structured dataset where each dissertation can be tagged with a binary ``Yes'' or ``No'' measure of its relation to each of the OSTP technology areas \citeApp{ostp2024critical}.

\noindent\textbf{Step 2: Technology subfield matching and filtering}

The top-level technology area classification is a first pass at our intended classification. It is also an intentionally broad filter. Several of the OSTP-defined domains---for example, \emph{Advanced Computing}, \emph{Advanced Engineered Materials}, \emph{Advanced Manufacturing}, and \emph{Biotechnologies}---create a broad catchment. The details of the OSTP guidance, however---including a listing of over 100 more detailed subfields within these technology areas---make clear that the technologies or applications its authors have in mind are often more specific.

To refine our measures, we implement a second-stage classification that further classifies each PhD graduate within a main technology area to each of its subfields---essentially re-applying our method for more specific target classes.\footnote{To prioritize precision in our classification, we exclude subfields that are ambiguous, duplicative, or overly generic. For example, under \emph{Advanced Computing}, we exclude ``Advanced modeling and simulation'' and ``Data processing and analysis techniques''. Although these terms reflect important computational practices, they are used broadly across many disciplines, making it difficult to reliably associate such labels with work situated specifically within advanced computing. Similarly, within \emph{Advanced and Networked Sensing and Signature Management}, we remove high-level application categories such as ``Health-sector sensing'', ``Energy-sector sensing'', ``Manufacturing-sector sensing'', etc.---all of which we think lack enough specificity to make useful distinctions. These refinements allow us to focus on subfields with clearer technical boundaries and greater specificity, supporting a classification that is more meaningful for the purposes of policy analysis and measurement of scientific labor.} This step helps us not only filter the initial classification to the specific subdomains that are considered ``critical'', but also to distinguish them: for example, this step distinguishes AI science related to ``Generative AI systems, multimodal and large language models'' from ``Technologies for improving AI safety, trust, security, and responsible use'', or biotech science related to ``Novel synthetic biology including nucleic acid, genome, epigenome, and protein synthesis and engineering'' from ``Biotic/abiotic interfaces.'' This second-stage classification is implemented by prompting \texttt{gpt-4o-mini} with the previously generated one-sentence summary, along with the dissertation title, subject terms, and keywords. The model is asked to assess whether the work is relevant to each subfield within a given technology area.

\noindent\textbf{Classification Results}

Across all 491,896 PhD graduates in our sample between 2000 and 2019, 206,872 (42.1\%) classify to at least one OSTP critical technology area through this procedure.

\begin{figure}[htbp]
\centering
\includegraphics[height=4in]{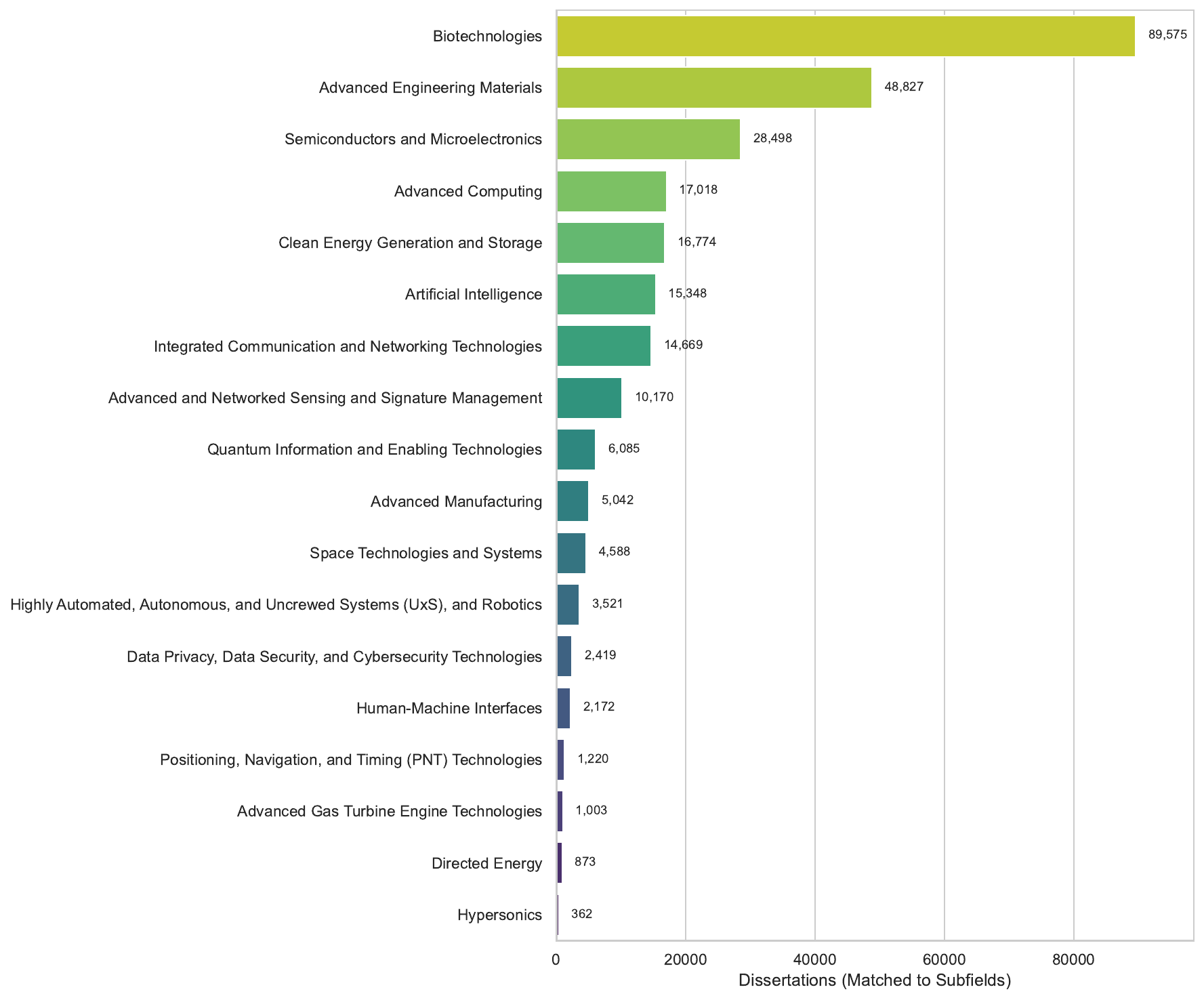}
\caption{\textbf{Total number of graduates by critical technology area, 2000-2019.} Figure shows the total number of US natural science and engineering PhD dissertations between 2000 and 2019 classified to each of 18 OSTP-defined critical technology areas, after applying both the first- and second-stage filters. See text for explanation of methodology.}
\label{fig:crittech_bytech_total}
\end{figure}

\begin{figure}[htbp]
\centering
\includegraphics[height=4in]{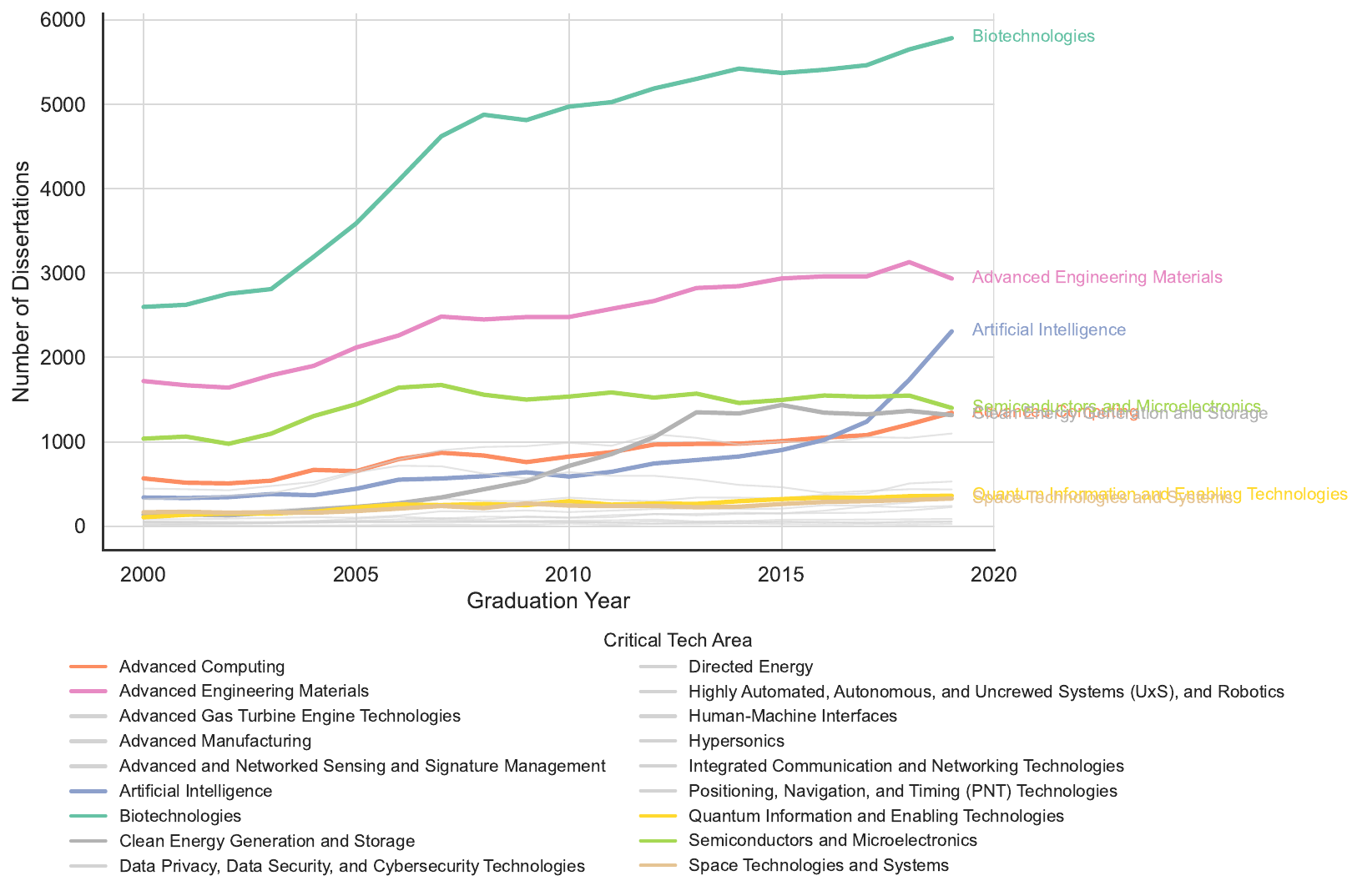}
\caption{\textbf{Annual number of graduates by critical technology area, 2000-2019.} Figure shows the annual number of US natural science and engineering PhD dissertations between 2000 and 2019 classified to each of 18 OSTP-defined critical technology areas, after applying both the first- and second-stage filters. See text for explanation of methodology.}
\label{fig:crittech_bytech_annual}
\end{figure}

Figure \ref{fig:crittech_bytech_total} shows the total number of dissertations that pass the two-stage classification by technology area, and Figure \ref{fig:crittech_bytech_annual} shows the number over time. \emph{Biotechnologies} is the largest technology area by a significant margin, with roughly 90,000 matched PhD graduates, reflecting the size of the life sciences in the US more generally (Figure \ref{fig:pqdt_vs_sed___all_subjects} shows that Biological Sciences is the largest SED field of US PhD graduates, by a similarly wide margin). The second largest (and fastest growing) technology area is AI, which was ranked sixth as recently as 2015 but experienced an inflection point ca. 2017. Other large areas include \emph{Advanced Engineering Materials} and \emph{Semiconductors and Microelectronics}, both of which have been associated with sustained federal and private investment. At the other end of the distribution, technology areas like \emph{Directed Energy}, \emph{Hypersonics}, and \emph{Advanced Gas Turbine Engines} have far fewer associated graduates. This difference could reflect limited university research capacity, greater concentration in non-academic institutions (e.g., national labs or defense contractors), but most likely reflects their specificity.

\noindent\textbf{Validation: Evidence and limitations}

Validating the results of this classification is made difficult by the absence of any ground truth sample. Whether a given dissertation's science is related to a particular technology is a simultaneously broad and nuanced question, for which there is no clear rubric, and thus a question even experts may disagree on.\footnote{``Critical technology areas'' itself is not an objectively category or taxonomy---as reflected by the fact that different government offices have produced different definitions of critical technologies over time, and have different (but overlapping) assessments of what technologies should be considered critical even today, including with varying degrees of specificity. The concept has traditionally instead been used as a general framework for policy choices and guidance around R\&D, immigration, export control, and other implied domains.} Our procedure for doing so developed iteratively as we assessed the results of different rule-based and LLM-based screens and experimented with prompt language. As we did so, we (informally) evaluated the results, comparing titles and abstracts to the LLM classification. Though this experimentation was not scientific, it helped improve our understanding of the data and build some \emph{prima facie} confidence in the results.

To more systematically assess the validity of our LLM-based classification, we conduct a case study focused on dissertations matched to the \emph{Quantum Information and Enabling Technologies} area (henceforth, ``quantum science''), examining where these PhD graduates are trained. Quantum science offers a useful testbed due to its topical specificity and its relatively well-documented institutional footprint, research centers, and publication patterns.

We begin by identifying the universities with the most quantum science dissertations between 2000 and 2019. We then examine whether the top PhD-producing universities in quantum science are also home to known quantum computing research centers, institutes, or degree programs. Table \ref{tab:quantum_top20}, for example, shows that nearly every one of the top 20 universities for quantum science PhDs over the past 20 years currently has a quantum science center.  

\begin{table}
\centering
\caption{\textbf{OpenAlex ``quantum'' research topics + parent fields and subfields.} Table lists the top 20 universities by quantum science PhD graduates between 2000 and 2019, as measured by our LLM-based procedure. For each university we also list what we assess to be its (current) principal quantum science research center, based on publicly available information.}
\label{tab:quantum_top20}
\resizebox{1\columnwidth}{!}{
\begin{tabular}{llcc}
\hline
\hline
Institution & Quantum Research Center & Dissertations & Publications \\
\hline
Massachusetts Institute of Technology
  & MIT Center for Quantum Engineering (CQE)
  & 240
  & 7,363 \\
University of California, Berkeley
  & Berkeley Quantum Info. \& Comput. Center
  & 197
  & 6,462 \\
Stanford University
  & Stanford Q-FARM
  & 188
  & 5,611 \\
Harvard University
  & Harvard Quantum Initiative \& QSE
  & 160
  & 4,289 \\
University of Maryland, College Park
  & JQI \& QuICS (UMD/NIST)
  & 150
  & 5,074 \\
University of Illinois at Urbana–Champaign
  & Illinois Quantum Info. Sci. \& Tech Center
  & 149
  & 5,059 \\
University of California, Santa Barbara
  & UCSB Quantum Foundry
  & 146
  & 5,645 \\
Princeton University
  & Princeton Quantum Initiative
  & 134
  & 5,428 \\
University of Michigan
  & U-M Quantum Engineering Sci. \& Tech
  & 122
  & 4,586 \\
University of Colorado at Boulder
  & JILA \& CUBit Quantum Initiative
  & 118
  & 2,432 \\
University of Wisconsin, Madison
  & Wisconsin Quantum Institute
  & 100
  & 2,808 \\
California Institute of Technology
  & Institute for Quantum Info. and Matter
  & 95
  & 3,784 \\
Northwestern University
  & Inst for Quantum Info. Res. \& Engineering
  & 94
  & 3,151 \\
Yale University
  & Yale Quantum Institute
  & 94
  & 2,453 \\
University of Washington
  & UW QuantumX
  & 86
  & 2,444 \\
University of New Mexico
  & UNM Center for Quantum Info. and Control
  & 86
  & 1,938 \\
Purdue University
  & Purdue Quantum Sci. \& Eng. Inst.
  & 83
  & 2,695 \\
University of Rochester
  & Center for Coherence and Quantum Science
  & 76
  & 2,243 \\
Pennsylvania State University
  & Materials Research Institute
  & 68
  & 3,648 \\
Cornell University
  & Cornell Quantum Science \& Engineering
  & 64
  & 3,130 \\
\hline
\hline
\end{tabular}
}
\end{table}

Though affirming, this evidence is also limited by its lack of variation: given rapidly growing interest in quantum computing and its underlying science, it may just be that many universities now have quantum science centers. To distinguish universities with stronger and weaker quantum science programs or research environments, we shift our attention from this extensive margin to the intensive margin. To do so, we measure universities' volume of scientific publications in quantum science. To do so we use OpenAlex data, filtering by publications whose OpenAlex-defined topics include the keyword ``quantum'' (see Table \ref{tab:oa_quantum} for a full list).

\begin{figure}[htbp]
\centering
\includegraphics[height=3in, trim={0 0 0 4cm}, clip]{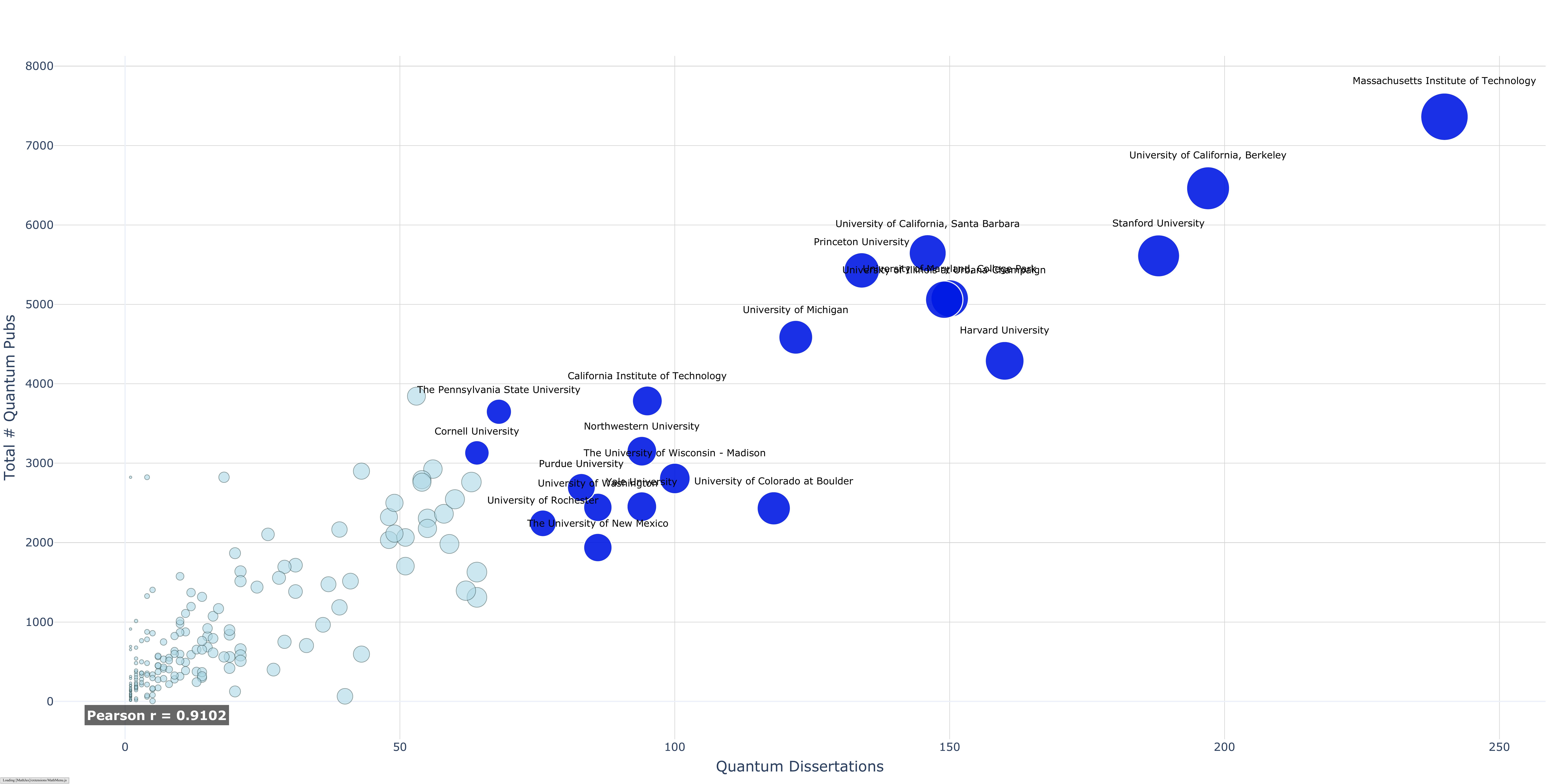}
\caption{\textbf{University quantum science publications vs. PhD graduates, 2000-2019.} Figure plots universities' number of quantum science publications between 2000 and 2019 (according to OpenAlex) against their number of quantum science PhD graduates (according to our LLM-based classification). Marker size proportional to an university's number of PhD graduates. The 20 universities with the most  graduates colored in blue and labeled.}
\label{fig:quantum_scatter}
\end{figure}

Using these data, Figure \ref{fig:quantum_scatter} plots the total number of quantum science publications at individual universities between 2000 and 2019 against the total number of quantum science PhD graduates over the same period. The top 20 universities from Table \ref{tab:quantum_top20} are colored in blue. The figure shows a very strong correlation between the two ($\rho$=0.9), indicating that the LLM-based procedure captures meaningful variation across universities and graduates in their relation to quantum science. Leading schools such as MIT, UC Berkeley, Stanford, Harvard, and the University of Maryland rank highly on both axes, lending face validity to the approach.

\begin{table}
\centering
\caption{\textbf{OpenAlex Top 20 US universities for quantum science, 2000-2019.} Table lists OpenAlex publication topics which include the word ``quantum''. According to OpenAlex, publications in its database are associated with ``Topics'' using an automated procedure that takes into account information about the work, including title, abstract, source (journal) name, and citations.}
\label{tab:oa_quantum}
\resizebox{1\columnwidth}{!}{
\begin{tabular}{lll}
\hline
\hline
Field & Subfield & Topic \\
\hline
Chemistry & Physical and Theoretical Chemistry & Chemical Reactions Involving Quantum Tunneling \\
Computer Science & Artificial Intelligence & Quantum Computing and Simulation \\
Computer Science & Artificial Intelligence & Quantum Information and Computation \\
Computer Science & Computational Theory and Mathematics & Design and Simulation of Quantum-dot Cellular Automata \\
Materials Science & Materials Chemistry & Applications of Quantum Dots in Nanotechnology \\
Materials Science & Materials Chemistry & Synthesis and Applications of Carbon Quantum Dots \\
Physics and Astronomy & Atomic and Molecular Physics, and Optics & Foundations of Electromagnetic Theory and Quantum Field Theory \\
Physics and Astronomy & Atomic and Molecular Physics, and Optics & Foundations of Quantum Mechanics and Interpretations \\
Physics and Astronomy & Atomic and Molecular Physics, and Optics & Parity-Time Symmetry in Optics and Quantum Mechanics \\
Physics and Astronomy & Atomic and Molecular Physics, and Optics & Quantum Coherence in Photosynthesis and Aqueous Systems \\
Physics and Astronomy & Atomic and Molecular Physics, and Optics & Quantum Dot Devices and Semiconductors \\
Physics and Astronomy & Atomic and Molecular Physics, and Optics & Quantum Effects in Helium Nanodroplets and Solids \\
Physics and Astronomy & Atomic and Molecular Physics, and Optics & Quantum Many-Body Systems and Entanglement Dynamics \\
Physics and Astronomy & Atomic and Molecular Physics, and Optics & Quantum Size Effects in Metallic Nanostructures \\
Physics and Astronomy & Atomic and Molecular Physics, and Optics & Semiconductor Spintronics and Quantum Computing \\
Physics and Astronomy & Atomic and Molecular Physics, and Optics & Slow Light Propagation and Quantum Memory \\
Physics and Astronomy & Condensed Matter Physics & Quantum Spin Liquids in Frustrated Magnets \\
Physics and Astronomy & Statistical and Nonlinear Physics & Cantorian-Fractal Theory of Quantum Physics \\
Physics and Astronomy & Statistical and Nonlinear Physics & Characterization of Chaotic Quantum Dynamics and Structures \\
Physics and Astronomy & Statistical and Nonlinear Physics & Characterization of Chaotic Quantum Dynamics and Structures \\
Physics and Astronomy & Statistical and Nonlinear Physics & Quantum Gravity and Noncommutative Field Theories \\
\hline
\hline
\end{tabular}
}
\end{table}

\subsubsection*{Tracking Graduates Over Time: OpenAlex}

To trace PQDT graduates' careers forward, we link dissertation records to scientific publications indexed by OpenAlex (a free, open-access bibliographic database cataloging global research; \citeApp{priem2022openalex}) using supervised deep learning. This approach offers significant advantages over traditional survey-based methods: we can observe career outcomes directly through publication records without relying on self-reporting, which often suffers from low response rates and potential bias (see discussion of survey measures later in the SM). Publications contain affiliation strings that allow us to infer geographic mobility and career trajectories over time by tracking where scientists publish from different institutions. However, the key challenge lies in correctly matching graduates to their subsequent publications, as names may change, be abbreviated differently, or be shared by multiple researchers. Our use of OpenAlex mechanically conditions the linked sample to PhD graduates who publish science, which we henceforth refer to as ``scientists''.

Our primary goal is to develop a high-fidelity dataset that accurately maps doctoral graduates to their subsequent research output. A central challenge in creating this mapping is name disambiguation---a problem which is particularly acute for graduates with common Asian-origin names. Given this paper's focus on foreign graduates and international migration, this population is also of particular analytical interest. To address this challenge, we develop a Deep Neural Network (DNN) model trained on a ground-truth dataset constructed from ORCID-linked records. This approach enables robust validation of matches and allows us to quantify link precision through standard performance metrics. The resulting linked dataset is the foundation for much of our analysis, identifying the location of researchers following their graduation. 

\paragraph{Training Data Construction}

To train our deep neural network for accurate name disambiguation, we require a large dataset of labeled examples--—both true matches between PQDT graduates and their publications, and false matches that represent common disambiguation errors. Creating such ground-truth data manually would be prohibitively expensive and time-consuming given the scale of our analysis. The availability of self-reported and authenticated ORCID records creates an opportunity to build a large sample of authenticated links between researchers and their publications, allowing us to automatically generate high-quality training labels at scale.

The training dataset was constructed by integrating PQDT dissertation records with ORCID profiles. The educational history available in ORCID profiles provides the key measures for linking ORCID profiles to their corresponding PQDT records. For PQDT records, we develop a custom name parser, adapted from the Python Human Name Parser library, to decompose author names into standardized components (first, middle, last) while removing special characters and inconsistencies.\footnote{See \texttt{\url{https://nameparser.readthedocs.io/en/latest/}}. Accessed March 2025.} We additionally extract and normalize publication years and institutional affiliations. We then apply similar pre-processing to ORCID data. After removing cases where an ORCID record matched to multiple doctoral graduates, the resulting PQDT-ORCID matched dataset comprises 71,563 PQDT doctoral graduates with their associated ORCID identifiers.

We then use these 71,563 ORCID identifiers to link PQDT records with publication data from OpenAlex. Doing so requires additional data cleaning and processing: although OpenAlex publications are often associated with an ORCID identifier, these identifiers are partially derived from OpenAlex's author disambiguation system---which is opaque, and which we determined to have many errors---such that OpenAlex ORCIDs may not always reflect authentic ORCID claims. To ensure ORCID identifiers are accurately attached to publications, we filter OpenAlex ORCID associations using authenticated ORCID claims from CrossRef's 2024 metadata dump. True matches between PQDT graduates and OpenAlex publications are then established when PQDT and OpenAlex records share the same authenticated ORCID identifier. False matches are identified as identical names associated with different authenticated ORCID identifiers. This approach provides high-quality labeled data necessary to train our deep learning model to distinguish between genuine and spurious links between PQDT graduates and OpenAlex publications.

We restrict the temporal scope of this linking to publications within 15 years of graduation to capture early-career research output. Our name matching criteria allow for variation in two ways: (i) matches can be based on shared first initials when complete middle names are available, and (ii) common nicknames (e.g., "Robert" and "Bob") are treated as equivalent.\footnote{Common nicknames from \texttt{\url{https://github.com/carltonnorthern/nicknames}}. Accessed March 2025.} When a match is based on first initials, we apply the Jaro-Winkler string similarity metric for validation, requiring a similarity greater than 90\%. This process generates a training dataset of 379,175 PQDT graduate-Open Alex publication links, comprising 51,132 true matches and 328,043 non-matches. The true matches correspond to 14,292 doctoral graduates in the PQDT dataset.

\begin{figure}[htbp]
   \centering
   \includegraphics[width=\textwidth]{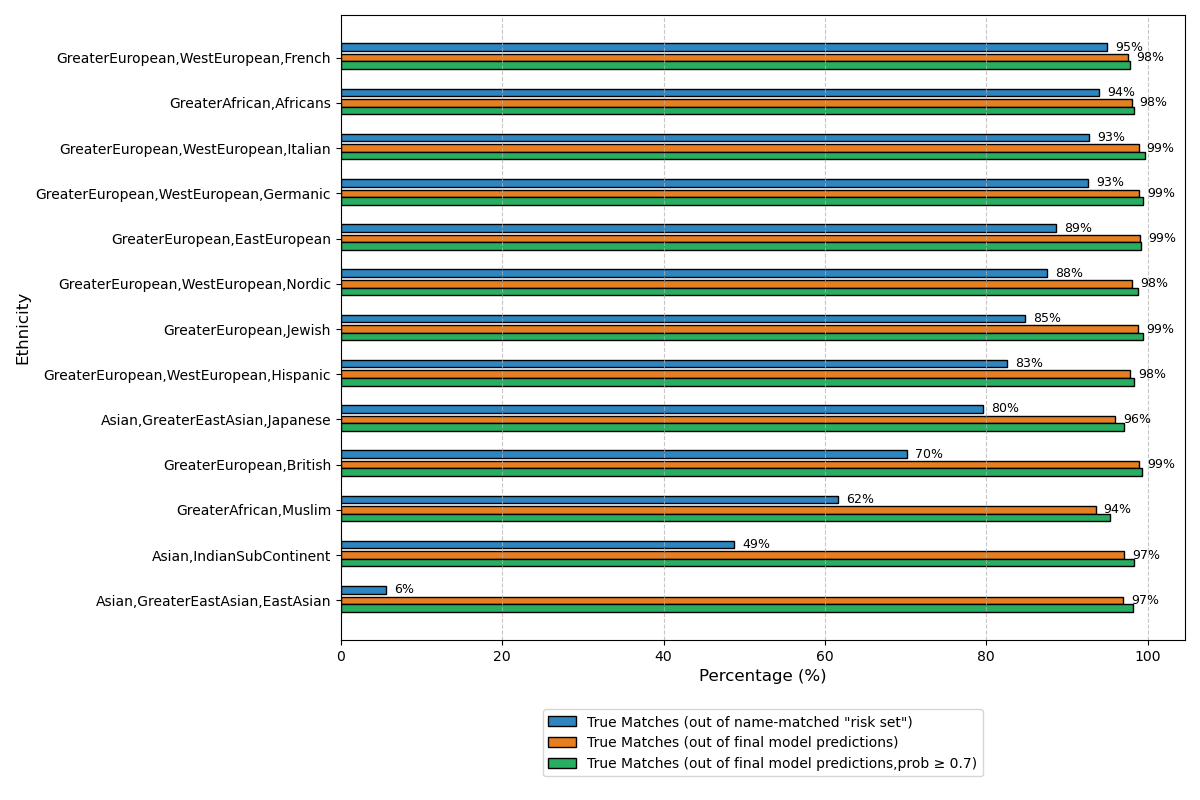}
   \caption[Training Dataset: Percentage of True Matches by Ethnicity]{\footnotesize \textbf{Percentage of True Matches by Ethnicity.} \linespread{1}\selectfont Figure shows three rates of true matches between PQDT graduates in our ORCID sample and OpenAlex publication authors, by name ethnicity. The blue bars show the rate of true matches among name-based candidate links (the ``risk set''). The orange bars show the rate of true matches among our final model's predictions, while the green bars show this rate for high-confidence predictions (match probability $\geq$ 0.7). Ethnicity estimated using raceBERT (\texttt{https://github.com/parasurama/raceBERT}).}
   \label{fig:ethnicity_pct_true}
\end{figure}

To assess variation in the rates of true matches for the training data, we process the sample of graduates using raceBERT, a Transformer-based model that predicts ethnicity from names.\footnote{See \texttt{\url{https://github.com/parasurama/raceBERT}}. Accessed March 2025.} The blue bars in Figure \ref{fig:ethnicity_pct_true} represent the percentage of true matches across ethnicities, highlighting substantial variation in match accuracy. Names predicted as East Asian show particularly low true match rates, with only about 6 percent of matches being correct. This likely reflects the high frequency and structural uniformity of East Asian names, which increase the likelihood of ambiguity during name-based identification. In contrast, names associated with West European ethnicities, such as French or Germanic, yield substantially higher true match rates, reflecting more distinctive naming conventions. African and Italian names also exhibit relatively high accuracy, possibly due to unique phonetic or morphological patterns. Nonetheless, name ambiguity remains a notable source of false matches across all groups.   These findings underscore the critical need for robust matching methods with transparent performance metrics to address disparities across diverse name structures, minimize false matches, and enhance the dataset's reliability. 

\paragraph{Model Features}

Our deep learning model incorporates multiple feature categories designed to capture the various dimensions of potential PQDT graduate-OpenAlex publication matches. Our approach leverages textual similarity, citation patterns, and name characteristics to create a multidimensional feature set for predicting matches.

\vspace{12pt}

\noindent \emph{Text-Based Features}

The first category of features we use represent semantic similarities between dissertations and publications. We generate text embeddings for dissertation and publication titles + abstracts (where available) using Google's \texttt{text\_embedding\_005} model, released through Vertex AI in November 2024. This model produces 768-dimensional vectors optimized for semantic similarity measurement. Using these embeddings, we compute several similarity metrics:

\begin{enumerate}

\item Euclidean distance between embedding vectors:
\begin{equation*}
    \text{euclidean}(\mathbf{d}, \mathbf{p}) = \sqrt{\sum_{i=1}^{n} (d_i - p_i)^2}
\end{equation*}

\item Manhattan distance between embedding vectors:
\begin{equation*}
    \text{manhattan}(\mathbf{d}, \mathbf{p}) = \sum_{i=1}^{n} |d_i - p_i|
\end{equation*}

\item Difference vector (element-wise subtraction of embeddings):
\begin{equation*}
    \text{diff}(\mathbf{d}, \mathbf{p}) = [d_1 - p_1, d_2 - p_2, \ldots, d_n - p_n]
\end{equation*}

\item Hadamard product (element-wise multiplication of embeddings):
\begin{equation*}
    \text{hadamard}(\mathbf{d}, \mathbf{p}) = [d_1 p_1, d_2 p_2, \ldots, d_n p_n]
\end{equation*}

\end{enumerate}

\vspace{12pt}

\noindent \emph{Citation Network Features}

We also construct features based on citation patterns that might indicate author identity:

\begin{enumerate}

    \item Direct self-citation: Binary indicator for whether the publication cites another paper where author name, institution, and publication year align with the dissertation record.
    
    \item Indirect self-citation: Binary indicator for whether the publication cites a paper that in turn cites another paper aligning with the dissertation record.
    
    \item Advisor/committee co-authorship: Binary indicator for co-authorship with dissertation advisor or committee members (based on name and institution).
    
    \item Advisor/committee citations: Binary indicator for citations to works by dissertation advisor or committee members (based on name and institution).
    
    \item Committee information availability: Binary indicators for presence of advisor and committee member names in the dissertation metadata.

\end{enumerate}

\vspace{12pt}

\noindent \emph{Additional Features}

We incorporate various additional features related to author names and variations:

\begin{enumerate}

    \item Jaro-Winkler distance between dissertation author name and OpenAlex display name.
    
    \item Jaro-Winkler distance between dissertation author name and OpenAlex raw name.
    
    \item Binary indicator for matching nicknames (e.g., ``Robert'' and ``Bob'').
    
    \item Name frequency: log occurrences of first, last, and full names in our PQDT sample.
      
    \item Predicted ethnicity derived from the raceBERT model.
    
    \item Temporal distance between dissertation and publication years.

\end{enumerate}

All features are inherently scaled appropriately for the neural network: similarity metrics and distances are bounded between -1 and 1, name frequencies are log-transformed, and the remaining features are binary indicators. Missing values in citation network features (e.g., due to incomplete committee information) are explicitly marked using binary indicators to allow the model to learn patterns associated with missing data.

\paragraph{Model Training and Selection}

We construct a balanced training dataset by randomly sampling 25,000 true matches and 25,000 non-matches from our ORCID-validated records. We consider several modeling approaches: a Deep Neural Network (DNN), Gradient Boosted Trees which capture feature interactions through tree ensembles, a Random Forest emphasizing robustness through bagging, and a Wide \& Deep architecture that combines linear models with neural networks to balance memorization of sparse features with generalization.

Table \ref{tab:model_performance} Panel A presents comparative performance metrics on this sample. The DNN achieves the highest performance across most metrics, with notably higher precision (0.966) and recall (0.971), leading to its selection as our primary linking model architecture.

\begin{table}[htbp]
\centering
\caption{\textbf{Comparison of Model Performance.} This table reports performance metrics for four record linkage approaches within our ORCID-based validation sample: Deep Neural Networks (DNN), Gradient Boosted Trees, Random Forests, and Wide \& Deep Learning. Panel (A) reports performance of each approach in first-stage linking, in balanced sample of 50,000 observations (25,000 true matches and 25,000 non-matches). Panel (B) reports performance in second-stage linking, using 20,000 observations from first stage predictions (10,000 true matches and 10,000 non-matches) and including first stage prediction probabilities as an additional feature. Sample restricted to publications within 15 years of dissertation completion.} \vspace{12pt}
\label{tab:model_performance}
\resizebox{0.85\columnwidth}{!}{
\begin{tabular}{lcccccc}
\hline
\hline
\multicolumn{7}{c}{Panel A: First Stage Models} \\
\hline
Model & Precision & Recall & Accuracy & F1 Score & Log Loss & ROC AUC \\
\hline
DNN & 0.966 & 0.971 & 0.969 & 0.969 & 0.090 & 0.995 \\
Boosted Tree & 0.957 & 0.963 & 0.960 & 0.960 & 0.113 & 0.992 \\
Wide \& Deep & 0.959 & 0.949 & 0.954 & 0.954 & 0.124 & 0.990 \\
Random Forest & 0.831 & 0.976 & 0.890 & 0.898 & 0.331 & 0.985 \\
\\
\hline
\multicolumn{7}{c}{Panel B: Second Stage Models} \\
\hline
Model & Precision & Recall & Accuracy & F1 Score & Log Loss & ROC AUC \\
\hline
DNN & 0.890 & 0.887 & 0.891 & 0.889 & 0.290 & 0.948 \\
Boosted Tree & 0.766 & 0.922 & 0.823 & 0.837 & 0.383 & 0.935 \\
Wide \& Deep & 0.885 & 0.887 & 0.888 & 0.886 & 0.290 & 0.946 \\
Random Forest & 0.841 & 0.908 & 0.870 & 0.873 & 0.384 & 0.942 \\
\hline
\hline
\end{tabular}}
\end{table}

The performance on this balanced training sample, however, may not fully translate to the complete dataset due to class imbalance in the full data. As Table \ref{tab:confusion_matrices}, Panel A shows, when applied to all 379,156 potential links, our first-stage model achieves a precision of 83.2\% (49,968 true positives out of 60,054 predicted links) and a recall of 97.7\% (49,968 true positives out of 51,132 true links). This degradation in precision from the balanced sample (96.6\%) to the full dataset (83.2\%) highlights the challenges posed by class imbalance, where even a small false positive rate can translate into a substantial number of incorrect links in absolute terms.

To address this challenge, we implement a two-stage cascading model approach. In the first stage, we apply our baseline model to the complete sample, identifying 60,054 potential matches. We then use this classified sample to train a second-stage model, constructing a balanced training dataset of 10,000 true matches and 10,000 false matches from the first-stage predictions. This second model incorporates the prediction probability from the first stage as an additional feature while using a simplified set of base features that excludes the textual similarity vector and Hadamard product to mitigate potential overfitting. As in the first stage, we evaluate several modeling approaches including gradient boosted trees, random forests, and neural networks.

As shown in Panel B of Table \ref{tab:model_performance}, the DNN model again achieved superior performance in the second stage, with the highest precision (0.890) and balanced performance across all metrics compared to alternative approaches. While the Boosted Tree achieved higher recall (0.922), this came at the cost of substantially lower precision (0.766), making it less suitable for our objective of minimizing false positives. The DNN's more balanced performance profile, with an F1 score of 0.889 and ROC AUC of 0.948, led to its selection as our second-stage model.

\begin{table}[htbp]
\centering
\caption{\textbf{Classification Results.} This table reports candidate PQDT-OpenAlex link classification results from the DNN model within our ORCID-based validation sample. First-stage linking is applied to the complete ORCID-validated sample (379,156 candidate links). Second-stage linking is applied to predicted matches from first stage (60,054 observations).}
\vspace{12pt}
\label{tab:confusion_matrices}
\resizebox{0.85\columnwidth}{!}{
\begin{tabular}{lrrrrrr}
\hline
\hline
& \multicolumn{3}{c}{Panel A: First Stage} & \multicolumn{3}{c}{Panel B: Second Stage} \\
\cline{2-4} \cline{5-7}
& Predicted & Predicted & & Predicted & Predicted & \\
True Status & Non-Match & Match & Total & Non-Match & Match & Total \\
\hline
Non-Match & 317,938 & 10,086 & 328,024 & 8,686 & 1,400 & 10,086 \\
Match & 1,164 & 49,968 & 51,132 & 5,922 & 44,046 & 49,968 \\
\hline
Total & 319,102 & 60,054 & 379,156 & 14,608 & 45,446 & 60,054 \\
\hline
\hline
\end{tabular}}
\end{table}

The classification results presented in Table \ref{tab:confusion_matrices} and visualized by the orange and green bars in Figure \ref{fig:ethnicity_pct_true} highlight the effectiveness of our two-stage approach. The first stage (Panel A) identified 60,054 potential matches from the full sample of 379,156 observations, with 49,968 true positives and 10,086 false positives. The second stage model (Panel B) successfully filtered out most of these false positives, reducing the total predicted matches to 45,446, of which 44,046 were true matches and only 1,400 were false matches. This second stage classifier thus significantly improves the quality of our matches by eliminating approximately 90\% of false positives while only sacrificing 10\% of true positives. Our final two-stage approach achieves an overall precision of 96.9\% (44,046 true positives out of 45,446 predicted matches) and recall of 86.1\% (44,046 out of 51,132 true matches). Importantly, as Figure \ref{fig:ethnicity_pct_true} shows, this precision is maintained across all ethnic groups, even in cases where the overall true match rate is lower due to inherent name ambiguity. Given that our primary goal is to maintain high precision in author-paper matches---which will be the basis for determining a PhD graduate's location and scientific productivity later in their career---we consider this tradeoff between precision and recall attractive for our application.

\begin{figure}[htbp]
    \centering
    \includegraphics[width=0.9\textwidth]{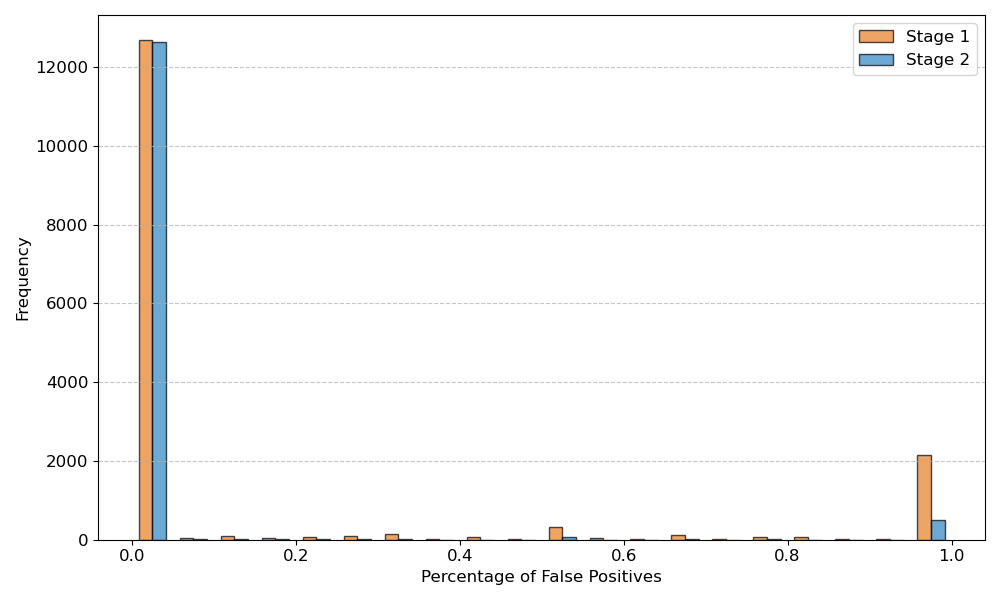}
    \caption{\textbf{Individual-level False Positive Rates by Stage.} Figure shows a histogram of individual-level false positive rates across graduates in our test sample. Each graduate's rate represents the proportion of their matches that are incorrect. Out of 19,797 graduates in the initial sample, 16,213 graduates are linked to at least one publication in the first stage, and 13,380 remain in the second stage. The second-stage model substantially reduces false positive rates, with most graduates shifting from higher to lower false positive categories and a portion dropping out from the sample.}
    \label{fig:false_positives_overall}
\end{figure}

\begin{figure}[htbp]
    \centering
    \includegraphics[width=0.9\textwidth]{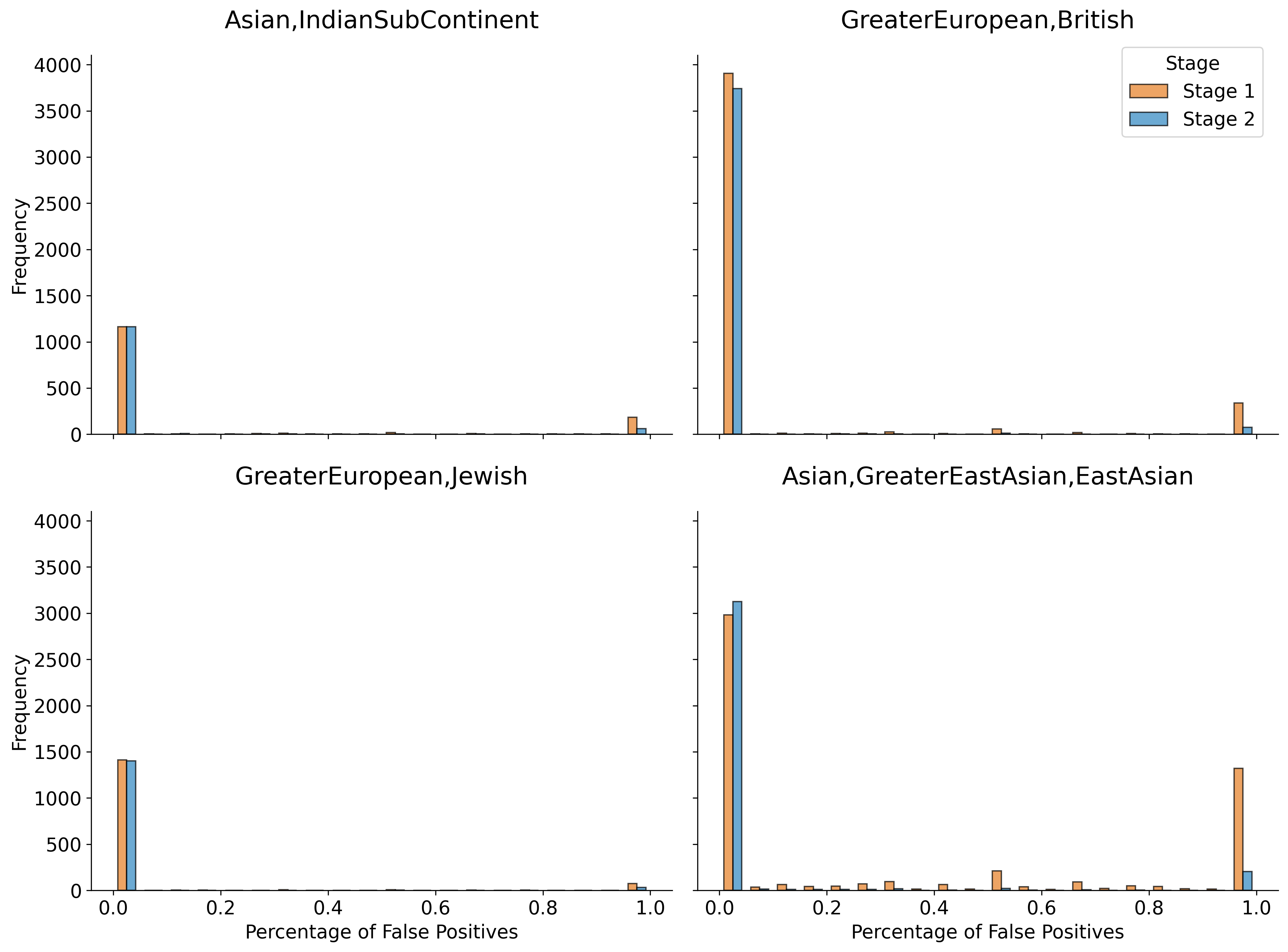}
    \caption{\textbf{Individual-level False Positive Rates by Stage and Ethnicity.} Figure shows histograms of individual-level false positive rates across graduates in our test sample, separated by ethnicity for the four largest ethnic groups. Orange bars show first-stage results, blue bars show second-stage results. Plotted data are a subset of those presented in Figure \ref{fig:false_positives_overall}.}
    \label{fig:false_positives_ethnicity} 
\end{figure}

Figures \ref{fig:false_positives_overall} and \ref{fig:false_positives_ethnicity} illustrate the distribution of false positives across the first and second stages of the model. In preparing these figures, we first calculate the false positive rate for each graduate in the linked sample as the proportion of that individual's matches that are incorrect, and we then plot a histogram of these individual-level rates across graduates in this sample.

Figure \ref{fig:false_positives_overall} shows the overall reduction in first-stage false positive rates (in orange) achieved by the second-stage classifier (in blue). The distribution reveals that most graduates have very low false positive rates, indicating nearly all their matches are correct. A small percentage of individuals have very high false positive rates, indicating of their most matches are incorrect. Only a small fraction of individuals appear in the intermediate range (with some true and some false matches). The second-stage model significantly shifts the distribution toward lower false positive rates, demonstrating a substantial decrease in incorrect matches at the individual level.

Figure \ref{fig:false_positives_ethnicity} illustrates how these improvements vary by ethnicity. In the first stage shown in orange, a notable number and proportion of false positives are concentrated among graduates with Asian names, reflecting challenges in disambiguating individuals in this population due to common/shared names. However, the second-stage model shown in blue addresses this issue, showing marked improvement in matching accuracy across ethnic groups.

\paragraph{Application to Full Sample}

We apply our two-stage predictive-model based linking approach to connect 1,246,495 graduates in PQDT between 1950 and 2022 to publications in OpenAlex. Using the same linking criteria employed in constructing our training data, we identify 271,824,338 potential links between graduates and publications. These potential links encompass 28,626,358 unique papers, with 1,106,602 graduates (88.8\% of the entire graduate sample) having at least one potential publication match that satisfies our basic name and timeline criteria.

Our first-stage model identifies 23,034,935 matches from this candidate set, representing approximately 8.5\% of all potential links. The second-stage classifier then filters these to 15,109,682 links. In a small number of cases, multiple graduates match to the same publication author. We disambiguate these by selecting the match with the higher match probability. Our final linked sample includes 14,997,158 links. Among these links, 977,368 unique doctoral graduates (78.4\% of all graduates) are linked to at least one publication, while 10,636,750 unique publications (37.1\% of candidate publications) are linked to at least one doctoral graduate. The average graduate in our linked sample is associated with 15.3 publications, and the average linked paper connects to 1.4 co-authors. The second-stage classifier's 45\% reduction (from 23 million to 15 million links) is consistent with the reduction observed in our ORCID-validated sample, providing additional confidence in the model's generalization to the full dataset. Moreover, the average of 15.3 publications per graduate is consistent with typical academic publication patterns.

\paragraph{Migration Indicators}

After establishing clean links between PQDT graduates and OpenAlex publications, we use these links to create indicators for graduates' out-migration. OpenAlex provides both raw author affiliation strings and disambiguated institutional data, including institution and country information. However, closer review suggests that OpenAlex's institution assignments are biased toward Western institutions and fail to label many foreign affiliations. We therefore process the raw affiliation strings independently to extract country information.

We implement a hybrid process to extract country information from author affiliation strings. First, we extract 5,808,404 unique raw affiliation strings from the OpenAlex data. We then process these strings using regular expression patterns matched against comprehensive lists of country names, US state names, and university names, which successfully resolves 3,939,436 affiliations (67.8\%) and links them to specific countries. For the remaining unresolved affiliations, we employ a large language model classification system using OpenAI's GPT-4o model with structured JSON output to ensure consistency in country code assignment. This LLM-based approach resolved an additional 906,605 observations, yielding a total resolution rate of 83.4\% across all unique affiliation strings. Expanding back to the full sample, these resolved affiliation strings are associated with 11,548,731 dissertation-publication links, which include 8,091,391 unique publications matched to 873,313 graduates (89\% of graduates with $\geq$1 linked publication). 

We construct several migration indicators based on publication affiliation patterns and validate these measures against employment histories from the ORCID subsample to identify the highest-performing measures. Our approach tracks graduates' annual publication output by country affiliation over a twenty-year window (five years pre-graduation to fifteen years post-graduation), applying varying confidence thresholds for publication-author matches (50\% baseline, 60\%, 70\%, and 80\%). We indicate potential migration events when a graduate's foreign-affiliated publications outnumber their US-affiliated publications in a given year. To filter out incorrect classifications arising from erroneous matches, temporary collaborations, or visiting positions, we test persistence requirements of varying lengths, creating separate indicators for migration events based on an author having one through four consecutive years of predominantly foreign affiliations. We measure timing of migration events as the onset of the longest such consecutive period, and we measure the destination country to be the modal foreign location in the onset year. By benchmarking these various indicator specifications against known employment transitions reported in our ORCID validation sample, we evaluate their relative precision in detecting actual geographic relocations and select the highest-performing measures for subsequent analysis.

Since our migration measures rely on observing multiple foreign affiliations to establish reliable patterns, we restrict our primary analysis to graduates with at least five matched publications. This threshold ensures sufficient publication histories to distinguish genuine relocations from noise in the matching process while maintaining a large sample of 528,434 graduates. As robustness checks, we examine how our results vary when applying more restrictive thresholds of at least 10 and 15 matching publications, yielding samples of 329,612 and 224,072 graduates, respectively. Results for graduates that meet these thresholds are provided later in the SM.

\begin{figure}[htbp]
\centering
\includegraphics[width=\textwidth]{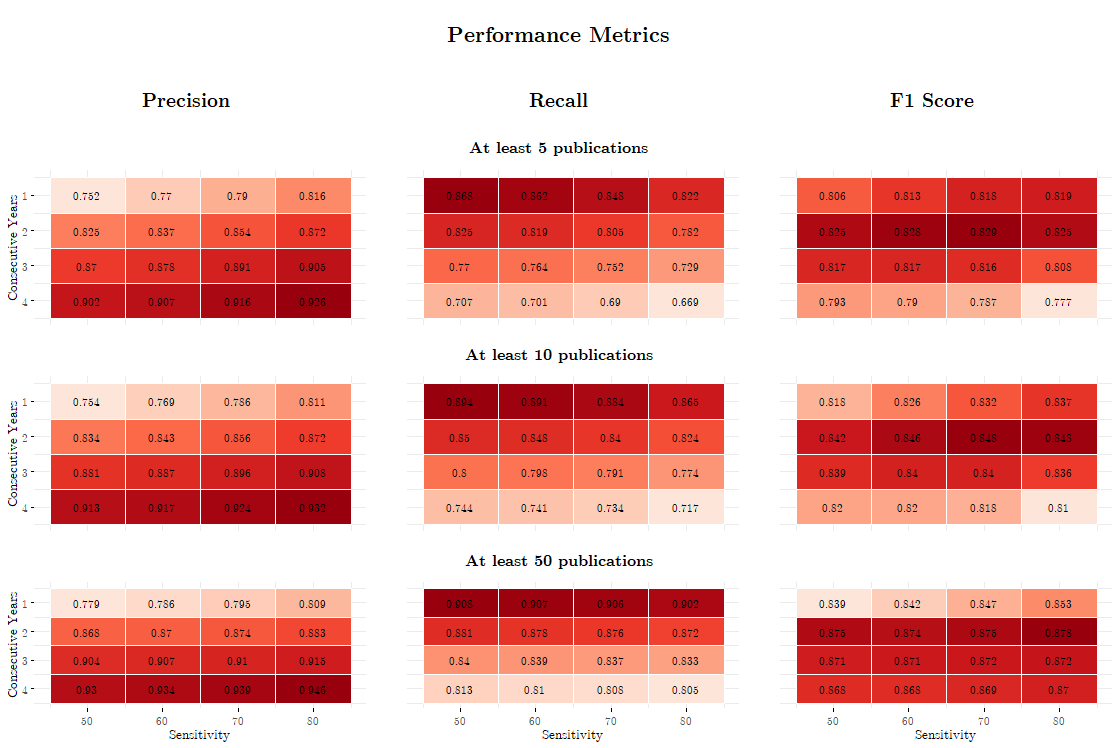}
\caption{\textbf{Migration Indicator Performance by Publication Threshold and Specification.} Heatmaps display precision, recall, and F1 scores for migration indicators across graduates with at least 5, 10, and 15 publications, validated against employment histories from our ORCID subsample. Each panel shows performance metrics for different combinations of match probability thresholds (50\%, 60\%, 70\%, 80\%) and consecutive year requirements (1-4 years).}
\label{fig:migration_indicator_performance}
\end{figure}

We evaluate the performance of our migration indicators using heatmaps that display precision, recall, and F1 scores across graduates with at least 5, 10, and 15 publications, validated against employment histories from our ORCID subsample of 63,697 graduates (47,461 of which have at least five matched publications). Each specification combines different dissertation-publication author match probability thresholds (50\%, 60\%, 70\%, 80\%; horizontal axis) with increasingly strict requirements on the number of consecutive years a graduate must be observed abroad to be measured as having migrated (1-4 years; vertical axis). Results are shown in Figure \ref{fig:migration_indicator_performance}. According to the heatmaps, the best-performing migration indicator sets the sensitivity threshold to 70\% and requires two consecutive years of foreign location, resulting in an F1 score of 0.829, precision of 0.854, and recall of 0.805. We use alternative parametrizations to establish confidence bounds for our analysis: for high precision, we apply an 80\% sensitivity threshold with four consecutive years (F1: 0.778, precision: 0.926, recall: 0.669), while for high recall, we use 50\% sensitivity with a single year requirement (F1: 0.806, precision: 0.752, recall: 0.868).

\begin{figure}[htbp]
\centering
\includegraphics[width=\textwidth]{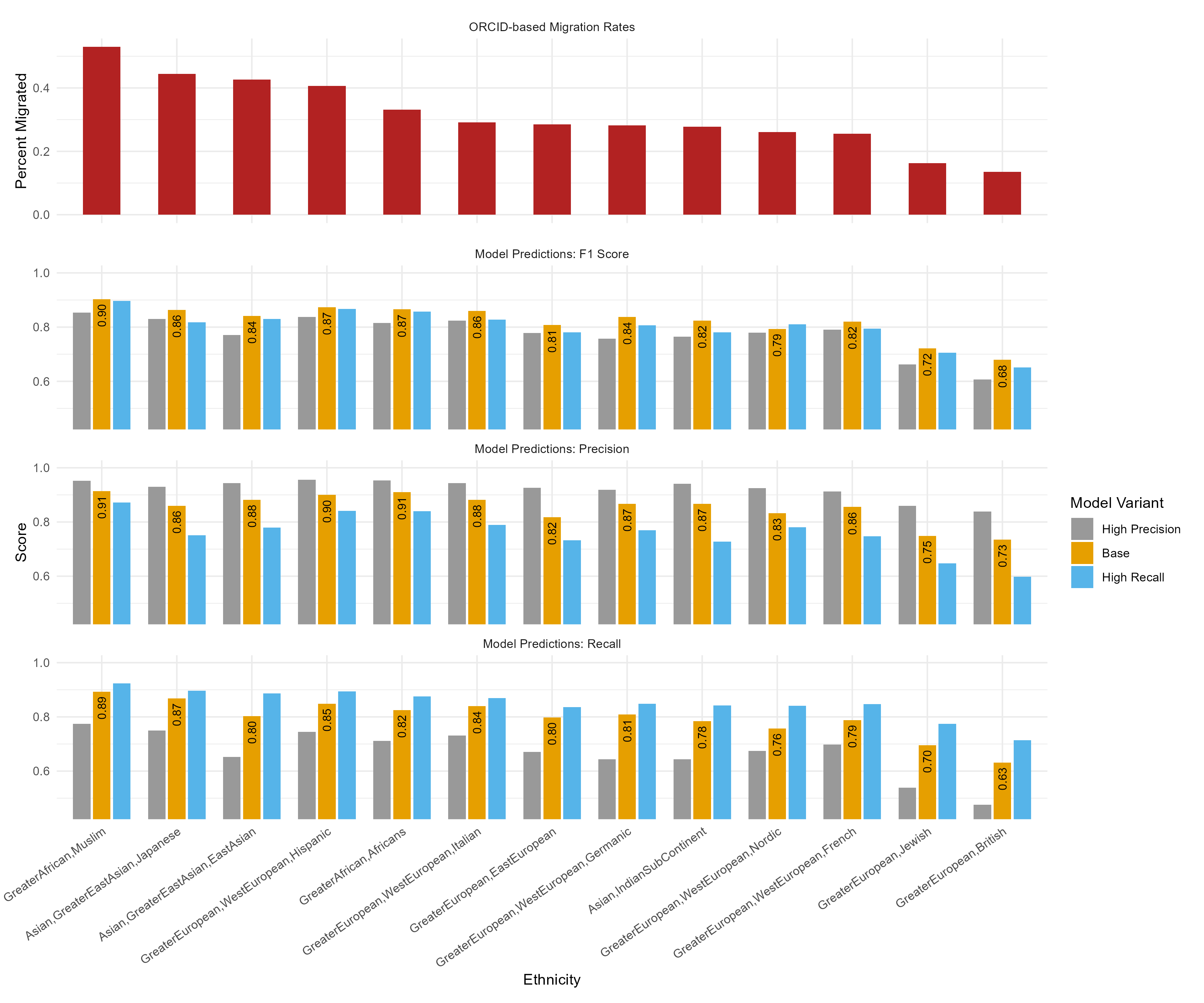}
\caption{\textbf{Migration Indicator Performance by Ethnicity.} Top panel shows ORCID-based migration rates across ethnic groups, ordered by decreasing migration propensity. Bottom panels display F1 scores, precision and recall for three model specifications (high precision, base, and high recall) validated against ORCID employment histories. Performance is strongest among ethnic groups with higher baseline migration rates.}
\label{fig:migration_validation_ethnicity}
\end{figure}

We further explore the performance of our indicators by examining variation in the model's performance across ethnic groups. Figure \ref{fig:migration_validation_ethnicity} presents performance metrics stratified by ethnicity and shows that indicator accuracy correlates positively with underlying migration propensities. Among populations with elevated international mobility---including graduates with East Asian, Muslim, and Indian names---our approach achieves F1 scores between 0.82 and 0.9. Performance somewhat deteriorates for ethnic groups characterized by lower migration rates, particularly those with British and European Jewish surnames. This performance gradient reflects the inherent challenge of detecting rare events, where limited positive cases reduce the model's ability to distinguish genuine migration signals from noise in publication affiliations.

\subsubsection*{Measuring PQDT Graduates' Scientific and Technological Impact}

We use three additional data sources to measure the impacts of migrating vs. non-migrating PhD graduates' impacts on science and technology. First, we use OpenAlex to measure academic citations accruing to graduates' scientific publications, which we interpret as a measure of scientific impact. Second, we use the publicly-available Reliance on Science dataset \citeApp{marx2020reliance, marx2022reliance} to measure patent citations to these graduates' scientific publications.\footnote{Reliance on Science is an open-access dataset that identifies non-patent literature cited by US and foreign (non-US) patents and links cited publications to OpenAlex identifiers.} In doing so, we include both front-page and in-text patent-paper citations, and only use those extracted by the dataset authors with a confidence score of 3 (out of 10) or higher, on the authors' recommendation. Third, we use publicly-available data on patent-paper pairs \citeApp{marx2024does} to identify patented inventions which emerged directly out of graduates' research. All three metrics are measured via publications, and the sample for which we have these measures will thus be mechanically restricted to PQDT graduates we can link to OpenAlex publications.

\subsubsection*{Additional Data Sources}

\paragraph{Survey of Earned Doctorates}

As a dissertation database, PQDT reports characteristics of PhD graduates' dissertations more so than the graduates themselves. More complete information on US PhD graduates is available from the NSF Survey of Earned Doctorates (SED).

The SED is an annual census of PhD students graduating from US universities, currently administered to nearly 50,000 graduates per year. The SED has been administered by NSF since 1958, collecting information on graduates' demographics, educational history, and immediate postgraduation plans.\footnote{For the most recent SED returns, and a description of the program, see \texttt{\url{https://ncses.nsf.gov/surveys/earned-doctorates/2023}}, accessed March 2025.}  The characteristics the SED reports have been relatively stable over time, though specific questions and response options sometimes vary. Importantly for our purposes, these questions include students' immigration status (native-born US citizen, naturalized US citizen, permanent resident, temporary resident) and citizenship country. 

We use SED data to measure the number of PhD graduates in our focal STEM fields who were non-US citizens at the time of graduation. Though the SED provides public-use tabulations of responses to some questions, the measures we are interested in are only available through the SED's microdata, which we accessed through a virtual data enclave hosted by NORC, under license from NCSES. We work with the 2019 copy of the Doctorate Records File (DRF), which compiles SED responses from 1958 to 2019. We first classify graduates to SED major fields and filter to our focal STEM fields. We then aggregate 2019 SED respondents by PhD completion year, citizenship status, and citizenship country. Using the resulting output we construct a time-series of foreign-citizen PhD graduates, which we use in Figure 1 of the paper. Note that because the DRF does not include names, we are unable to link it to our PQDT data (or other individually-identified measures) to measure PQDT graduates' nationality, place of birth, or other features.

\paragraph{Survey of Doctorate Recipients}

The Survey of Doctorate Recipients (SDR) is a separate data collection program administered by NCSES approximately every two years. The SDR seeks to collect information on a sample of past doctoral graduates of US institutions, including their location, field, employment status, occupation, industry, job function, and more. The SDR typically surveys roughly 10\% of known graduates, with response rates around 65\%.\footnote{For the most recent SDR returns, and a description of the program, see \texttt{\url{https://ncses.nsf.gov/surveys/doctorate-recipients/2023}}, accessed March 2025.}

In principle, the SDR could be used to evaluate migration patterns of US PhD graduates, and it has been used towards this end by prior research (e.g., Corrigan et al. 2022; \citeApp{corrigan2022long}). However, for our purposes in this paper, the SDR has several limitations. The first is the low sampling rate, which requires us to make population inferences from a 10\% sample. The second is incomplete response rate and the risk of nonrandom response---especially if the very graduates we are interested in (those who migrate) are disproportionately nonresponsive. Reflecting this challenge, Corrigan et al. note that NSF has explicitly advised against using some editions of the SDR to evaluate international mobility due to high nonresponse rates among PhD graduates living outside the US. The SDR also constrains us to studying survey variables, which precludes studying scientific productivity and impact of US PhD graduates, or linking them to critical technology areas. Finally, each SDR survey wave returns information on a cross-section of past PhD graduates and documents where they reside at a single moment in time, precluding the construction of longitudinal patterns for each cohort (e.g., its 5, 10, and 15-year emigration rate, as in Figure 1 of the paper). Given these limitations, we chose to forgo the SDR and instead develop new methods of linking PhD graduates forward to their scientific careers using publication and patent data.

\pagebreak


\subsection*{Supplementary Figures}

\begin{figure}[H]
\centering
\includegraphics[width=0.8\linewidth]{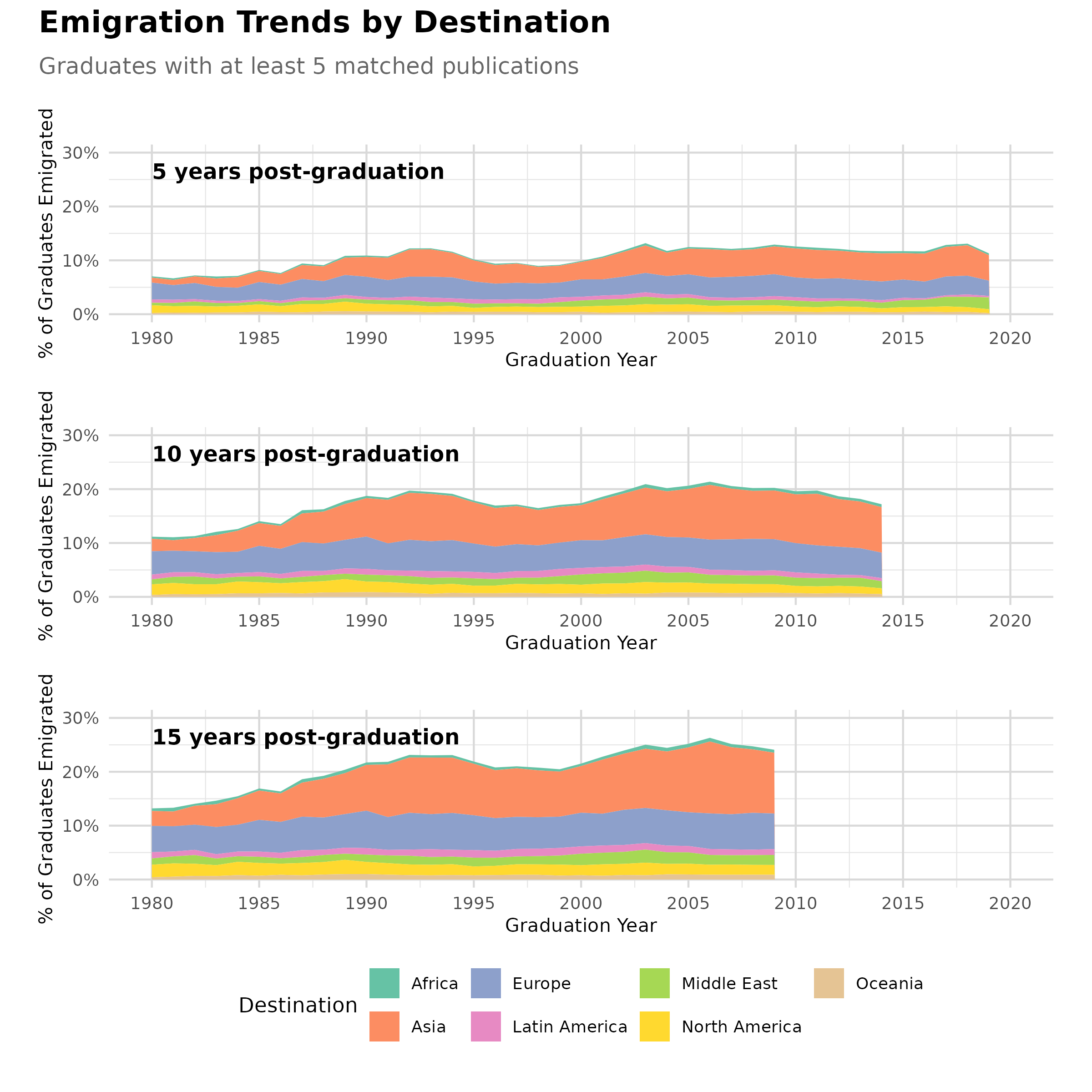} \\
\caption{\textbf{This figure shows destination regions over time of graduates we identify as having emigrated to a foreign country 5, 10, and 15 years post-PhD.}}
\label{fig:trends_region}
\end{figure}

\begin{figure}[H]
\centering
\includegraphics[width=0.8\linewidth]{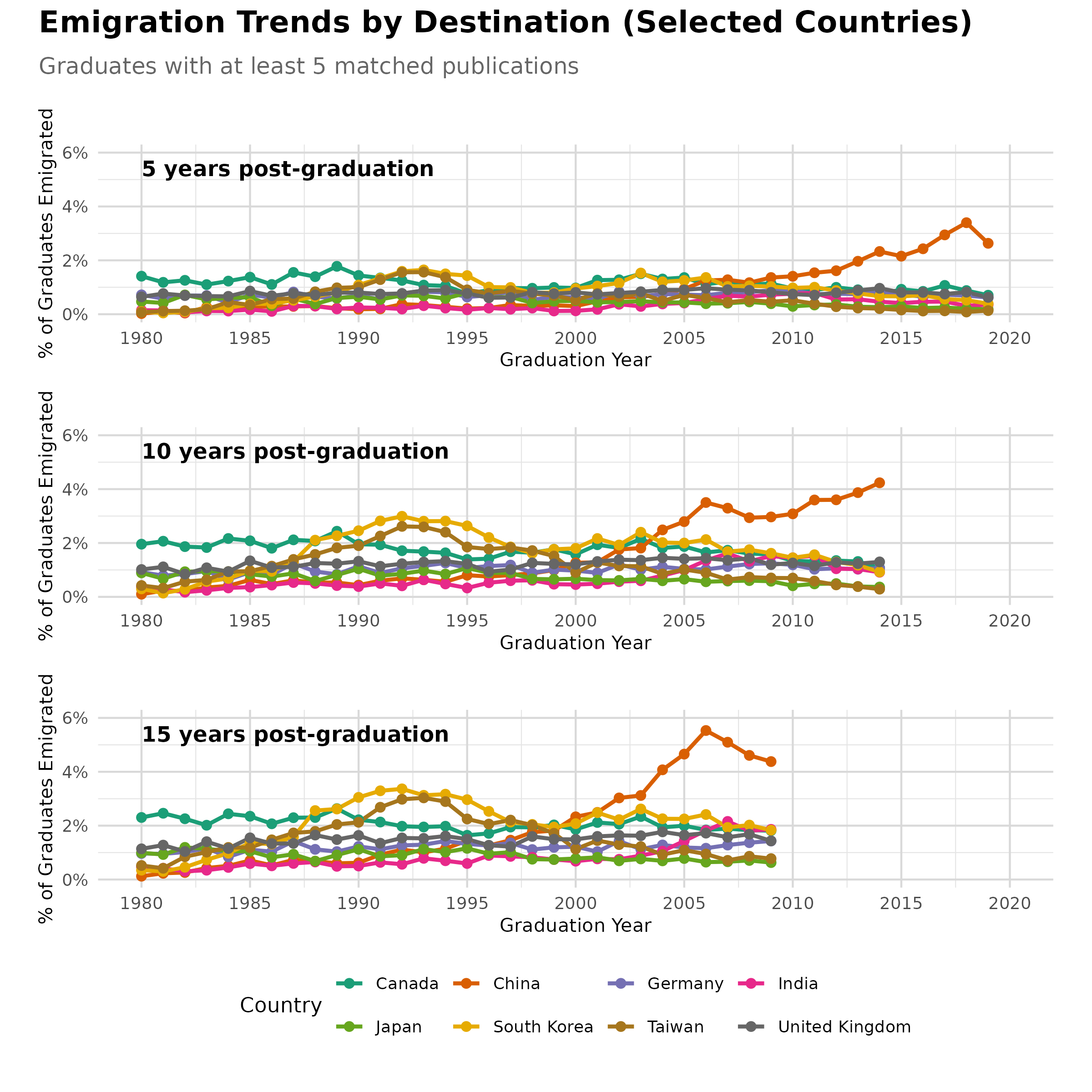} \\
\caption{\textbf{This figure shows destination countries over time of graduates we identify as having emigrated to a foreign country 5, 10, and 15 years post-PhD.}}
\label{fig:trends_country}
\end{figure}

\begin{figure}[H]
\centering
\includegraphics[width=0.8\linewidth]{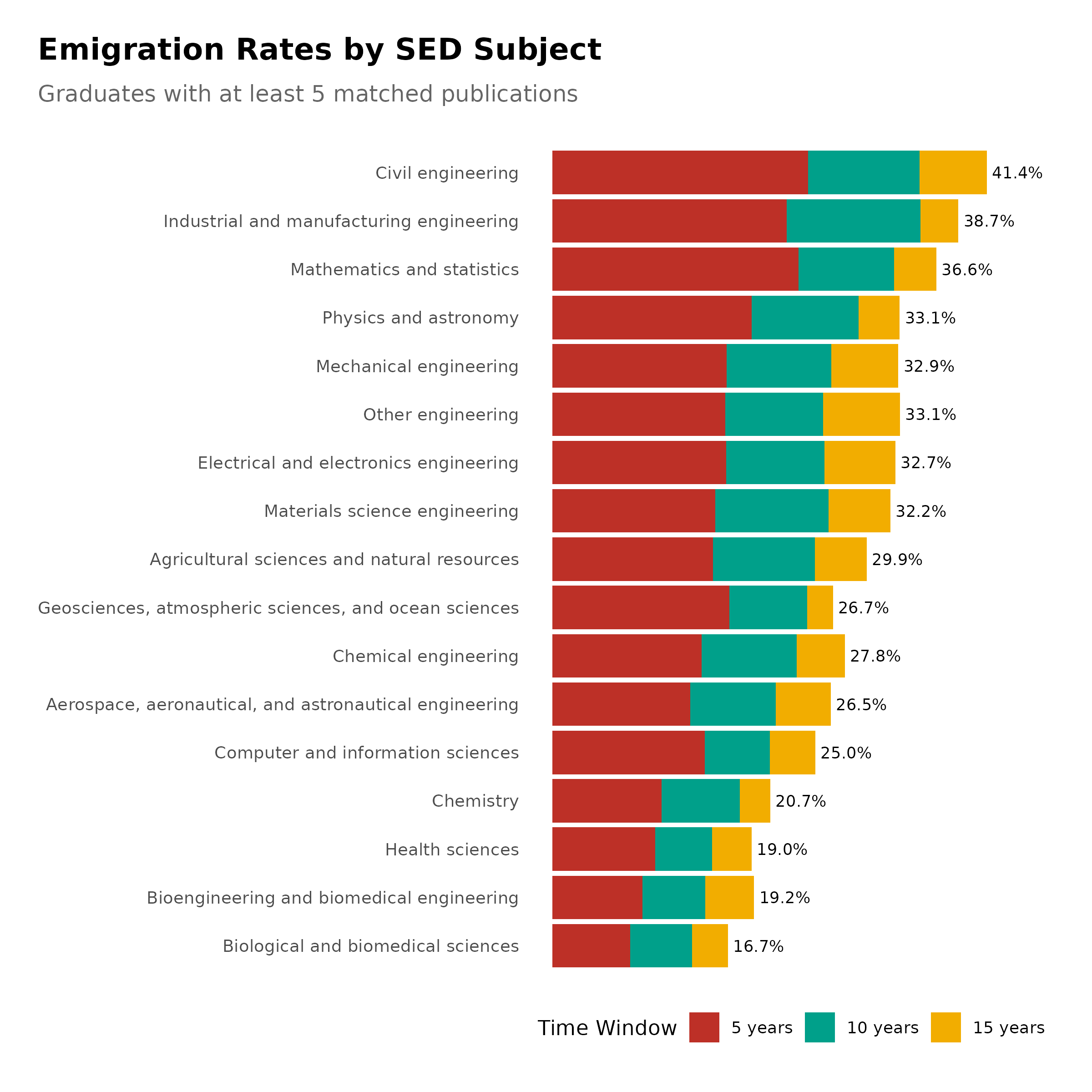} \\
\caption{\textbf{This figure disaggregates the results in Figure \ref{fig:figure1}C documenting emigration rates by broad field into more detailed major fields.}}
\label{fig:all_fields}
\end{figure}

\begin{figure}[H]
\centering
\includegraphics[width=0.8\linewidth]{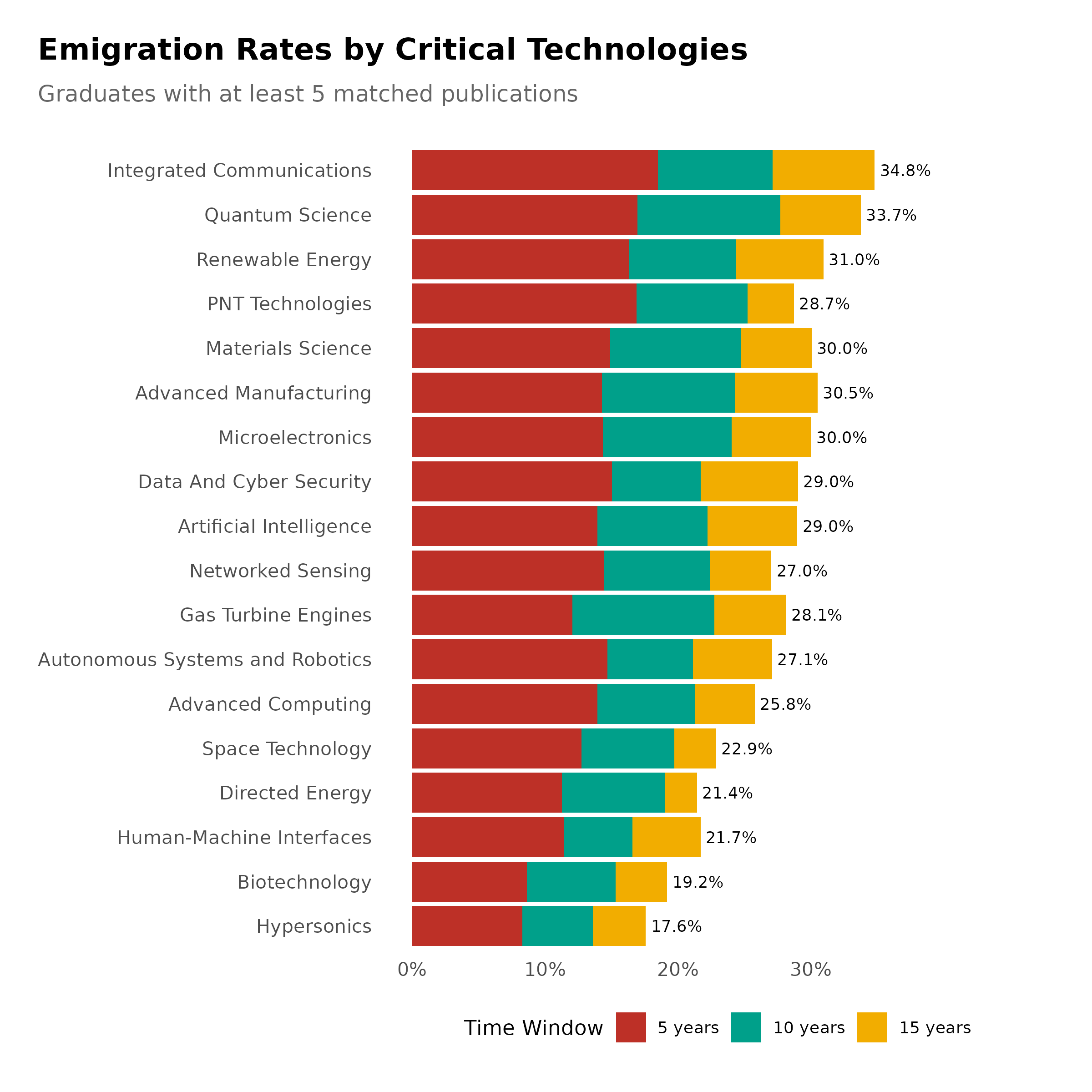} \\
\caption{\textbf{This figure extends the results in Figure \ref{fig:figure1}C documenting emigration rates by critical technology area to all OSTP technology areas.}}
\label{fig:all_techs}
\end{figure}

\begin{figure}[H]
\centering
\includegraphics[width=0.8\linewidth]{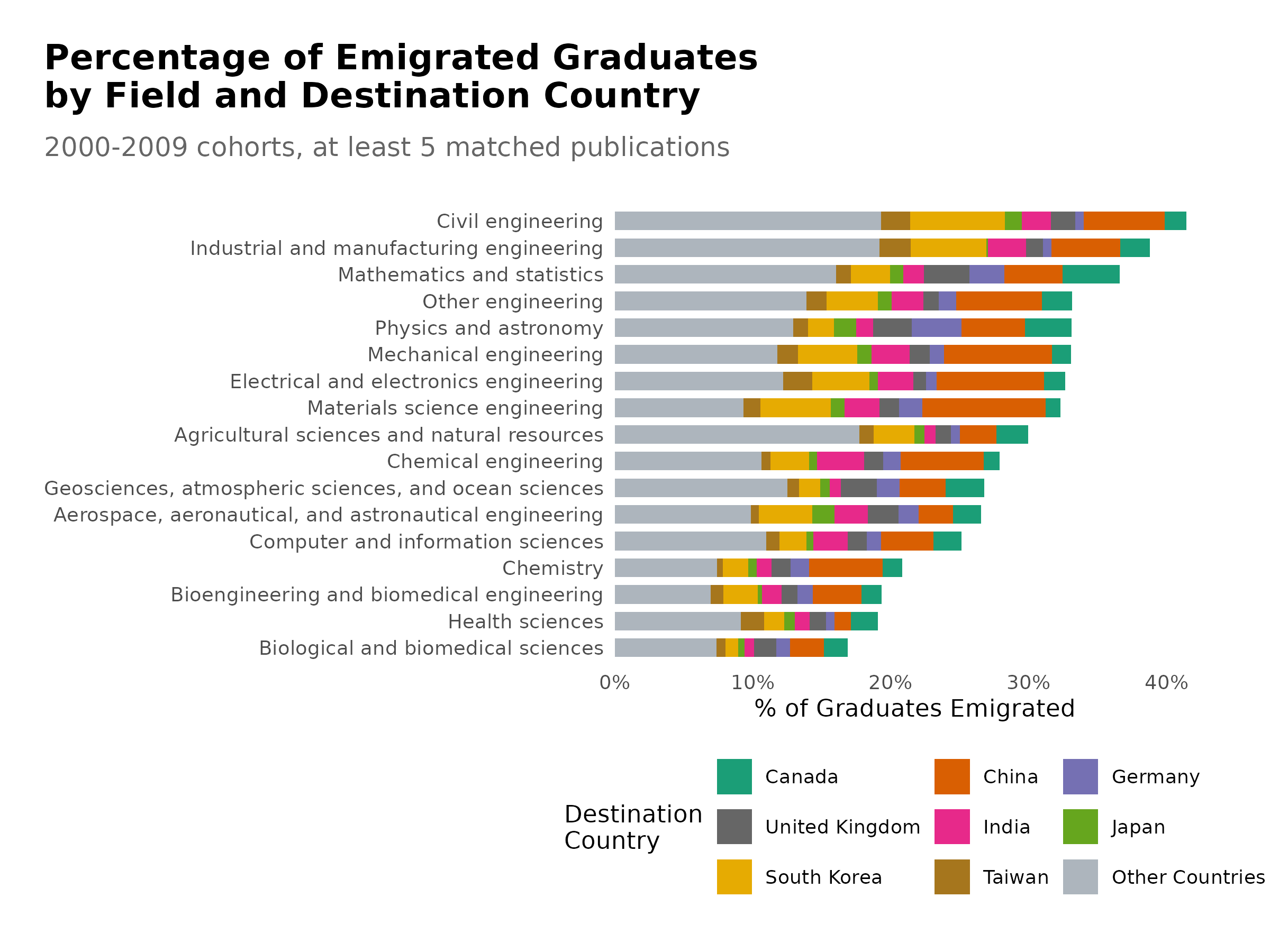} \\
\caption{\textbf{This figure shows the share of graduates in the 2000-2009 cohorts emigrating by major field (as in Figure \ref{fig:all_fields}), broken out by destination country. We limit the sample to graduates through 2009 to ensure we capture both short-run (5-year) and long-run (15-year) outcomes for all sampled individuals.}}
\label{fig:subject_country}
\end{figure}

\begin{figure}[H]
\centering
\includegraphics[width=0.8\linewidth]{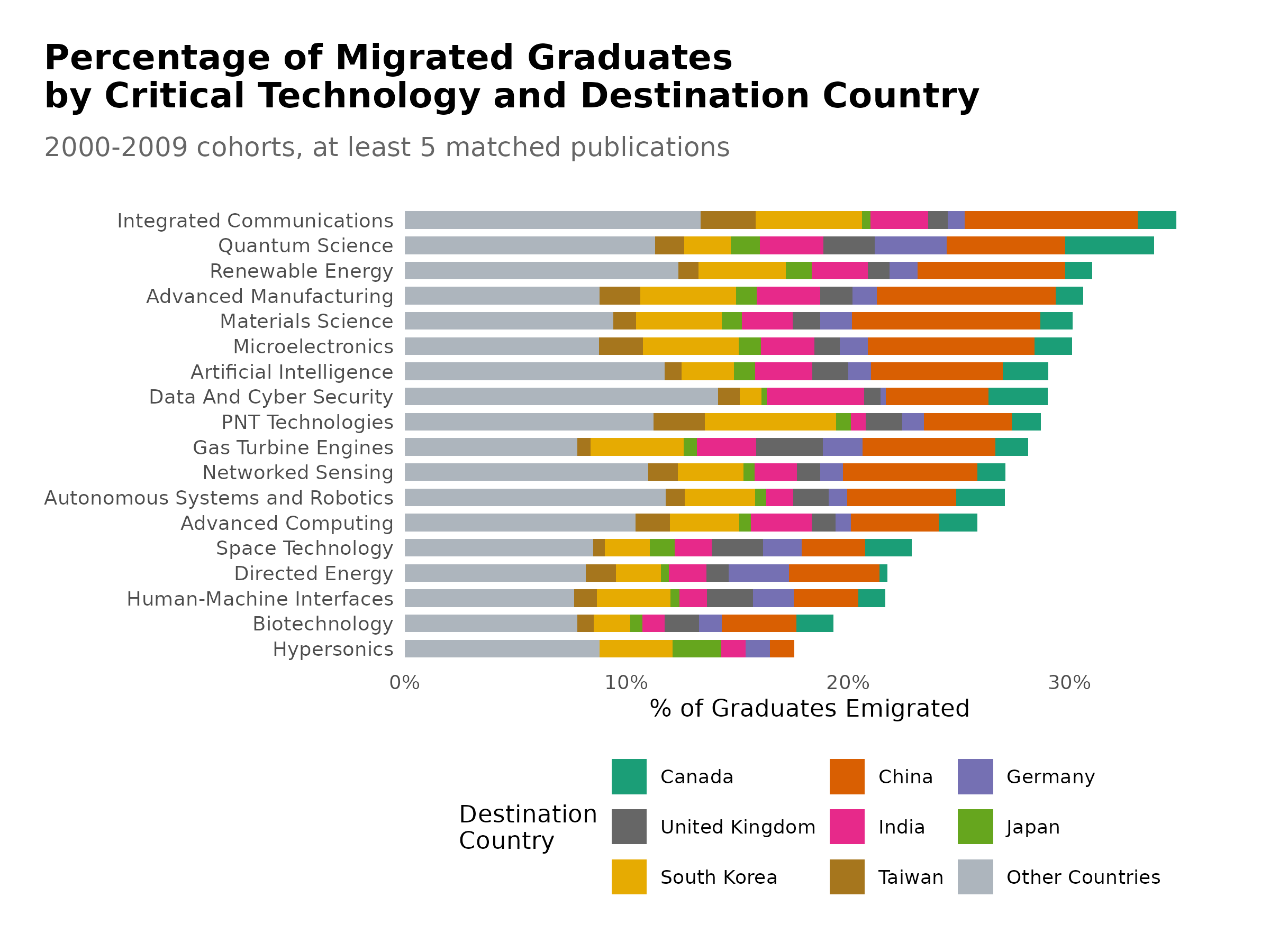} \\
\caption{\textbf{This figure shows the share of graduates in the 2000-2009 cohorts emigrating by technology (as in Figure \ref{fig:all_techs}), broken out by destination country. We limit the sample to graduates through 2009 to ensure we capture both short-run (5-year) and long-run (15-year) outcomes for all sampled individuals.}}
\label{fig:tech_country}
\end{figure}

This section presents several extensions and robustness checks on the results in Figures \ref{fig:figure1} and \ref{fig:figure2} of the paper. In Figure \ref{fig:trends_region}, we present 5/10/15-year emigration rates over time by destination region, and in Figure \ref{fig:trends_country} by destination country. Figure \ref{fig:all_fields} presents emigration rates by major field (disaggregating from broad fields shown in Figure \ref{fig:figure1}C), and Figure \ref{fig:all_techs} by technology, for all OSTP critical and emerging technology areas (expanding on the subset presented in Figure \ref{fig:figure1}C). Figure \ref{fig:subject_country} shows the principal destination countries of emigrating graduates in each major field, and Figure \ref{fig:tech_country} does so for graduates in each critical technology area.

\begin{figure}[H]
\centering
\includegraphics[width=0.8\linewidth]{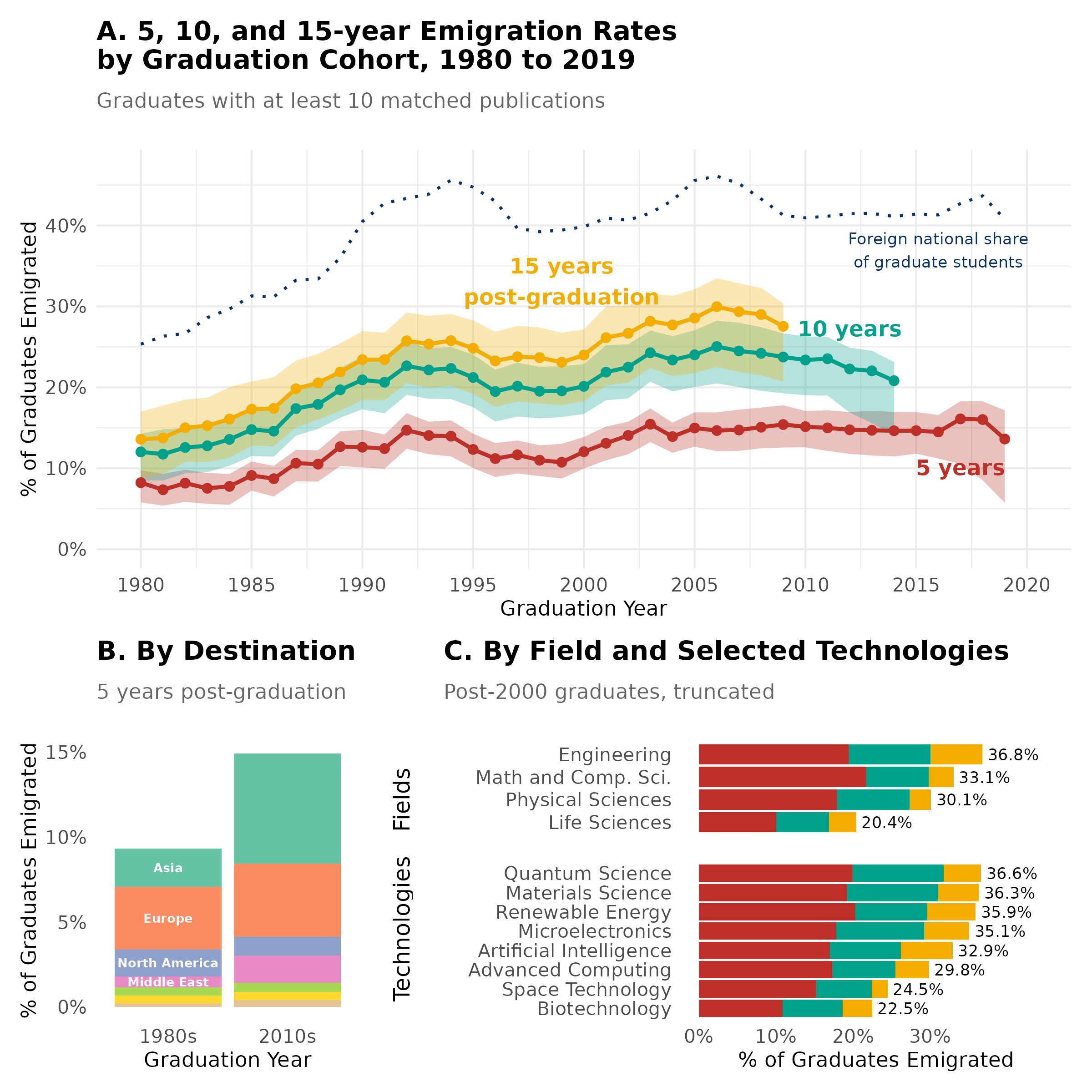} \\
\caption{\textbf{This figure presents a variant of Figure \ref{fig:figure1}, conditioning the sample to graduates with at least 10 linked publications.} Panel A shows the foreign national share of STEM PhD graduates since 1980 (black dotted line, based on the Survey of Earned Doctorates) and share of all graduates we identify as publishing in a foreign country 5, 10, and 15 years post-PhD (red, orange, and yellow lines, respectively, based on our PQDT graduate sample linked to future publication affiliations). Colored bounds represent confidence intervals by varying model specifications. Panel B shows the share emigrating to the specified global regions after 5 years in the 1980s vs. 2010s. Panel C shows the share emigrating by broad field and in select technology areas. All charts condition on graduates with at least 5 linked publications.}
\label{fig:figure1_10pubs}
\end{figure}

\begin{figure}[H]
\centering
\includegraphics[width=0.8\linewidth]{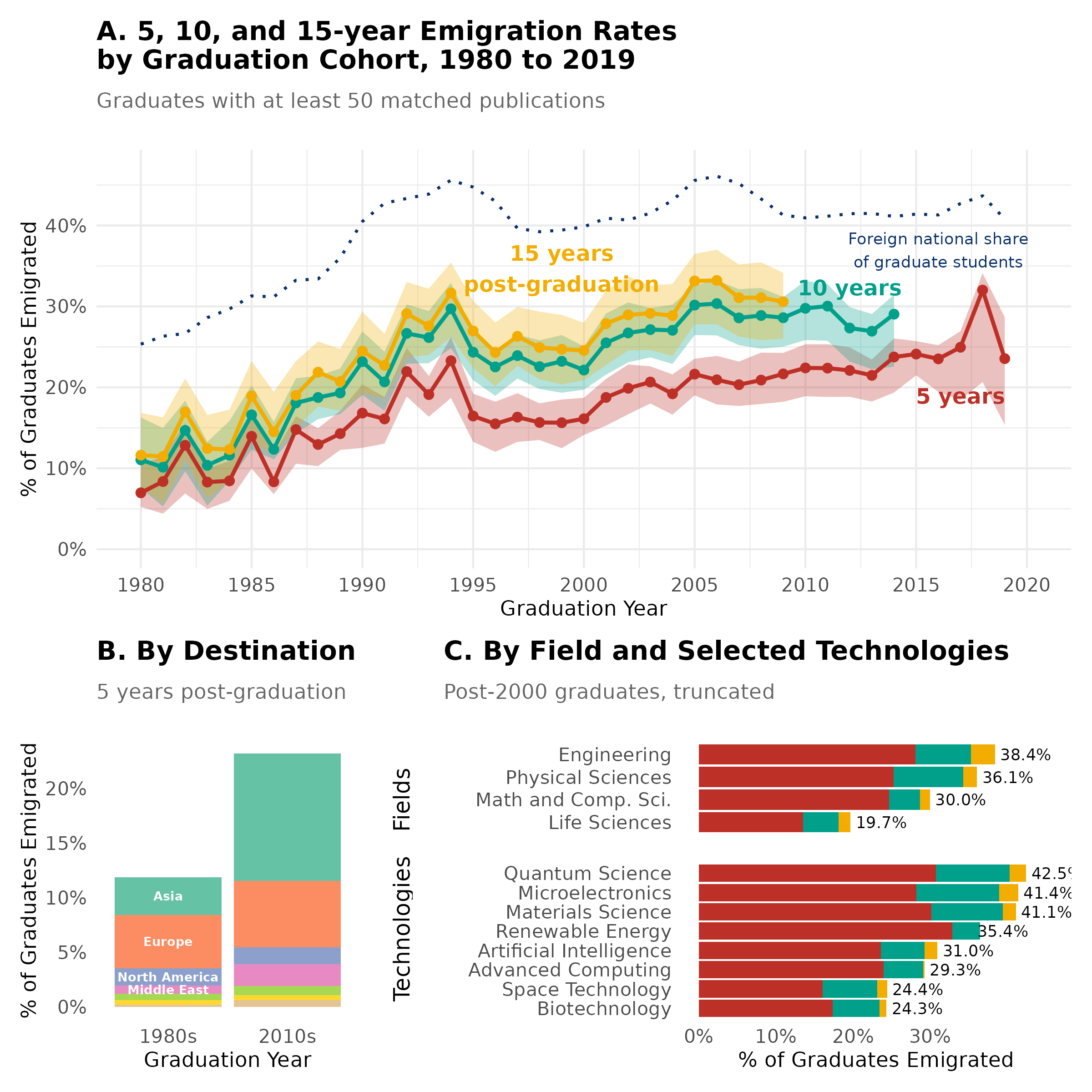} \\
\caption{\textbf{This figure presents a variant of Figure \ref{fig:figure1}, conditioning the sample to graduates with at least 50 linked publications.} Panel A shows the foreign national share of STEM PhD graduates since 1980 (black dotted line, based on the Survey of Earned Doctorates) and share of all graduates we identify as publishing in a foreign country 5, 10, and 15 years post-PhD (red, orange, and yellow lines, respectively, based on our PQDT graduate sample linked to future publication affiliations). Colored bounds represent confidence intervals by varying model specifications. Panel B shows the share emigrating to the specified global regions after 5 years in the 1980s vs. 2010s. Panel C shows the share emigrating by broad field and in select technology areas. All charts condition on graduates with at least 5 linked publications.}
\label{fig:figure1_50pubs}
\end{figure}

We then re-examine our main results under different variable definitions and model parametrizations. Figures \ref{fig:figure1_10pubs} and \ref{fig:figure1_50pubs} reproduce Figure \ref{fig:figure1} restricting to graduates with at least 10 and 50 linked publications, respectively (vs. a minimum of 5 in Figure \ref{fig:figure1}). Patterns are directionally similar, though our sample shrinks considerably when we filter to graduates with more publications (relatively few graduates accumulate 50 publications in their first 15 postgraduate years), and estimated emigration rates grow; this is in part attributable to this high-publication subsample being more susceptible to there being one or more bad links which may lead us to misclassify a non-emigrant as an emigrant. We consider our results based on the 5-publication threshold most reliable.

\begin{figure}[H]
\centering
\includegraphics[width=0.75\linewidth]{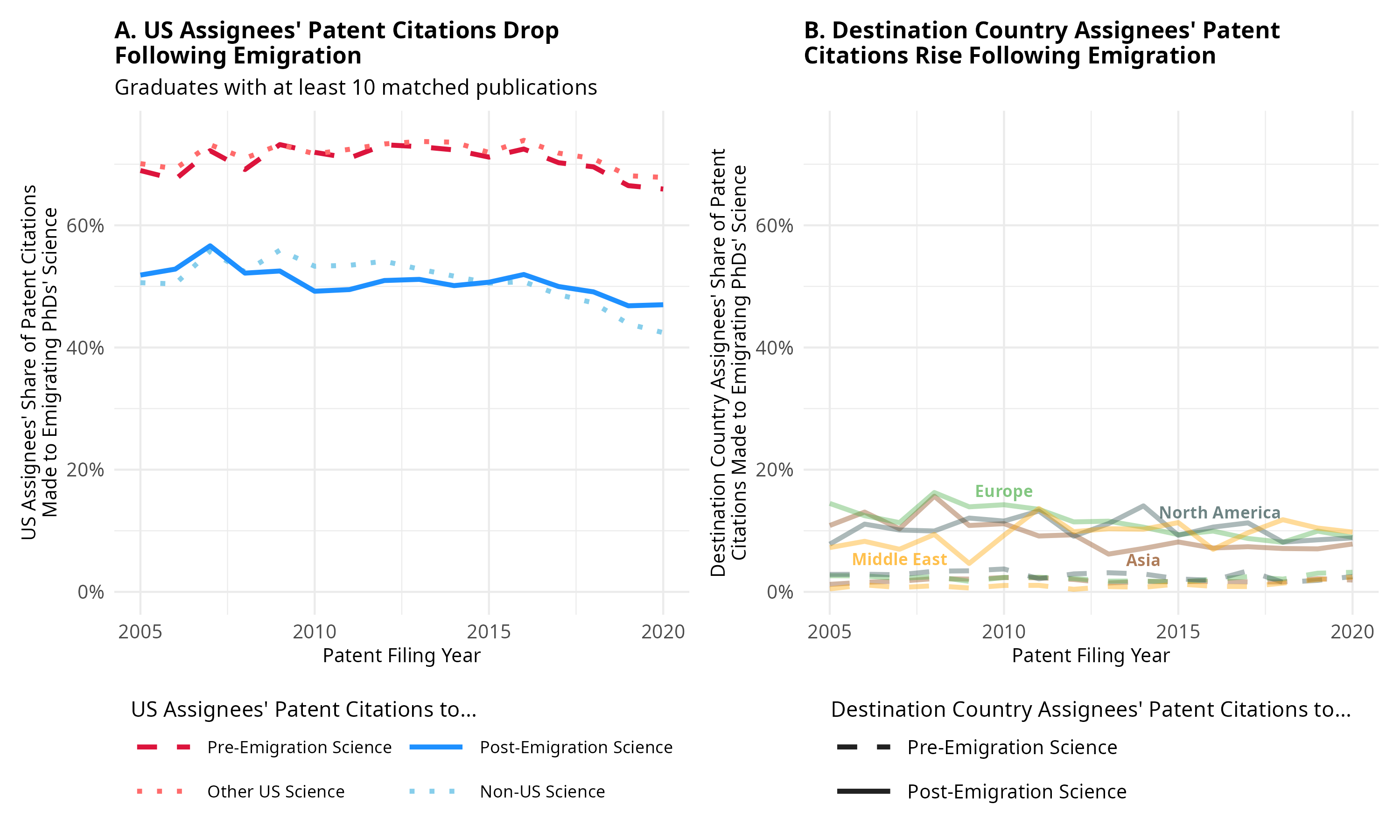} \\
\caption{\textbf{This figure presents a variant of Figure \ref{fig:figure2}, conditioning the sample to graduates with at least 10 linked publications.} Panel A evaluates US assignees' share of citations made each year in global patents to PhD graduates' science published before vs. after they leave the US (dashed red and solid blue lines, respectively). For comparison, the figure also provides US assignees' share of citations to non-leaver graduates' science and non-US science (dashed blue and red lines). Panel B shows a countervailing effect in destination countries.}
\label{fig:figure2_10pubs}
\end{figure}

\begin{figure}[H]
\centering
\includegraphics[width=0.75\linewidth]{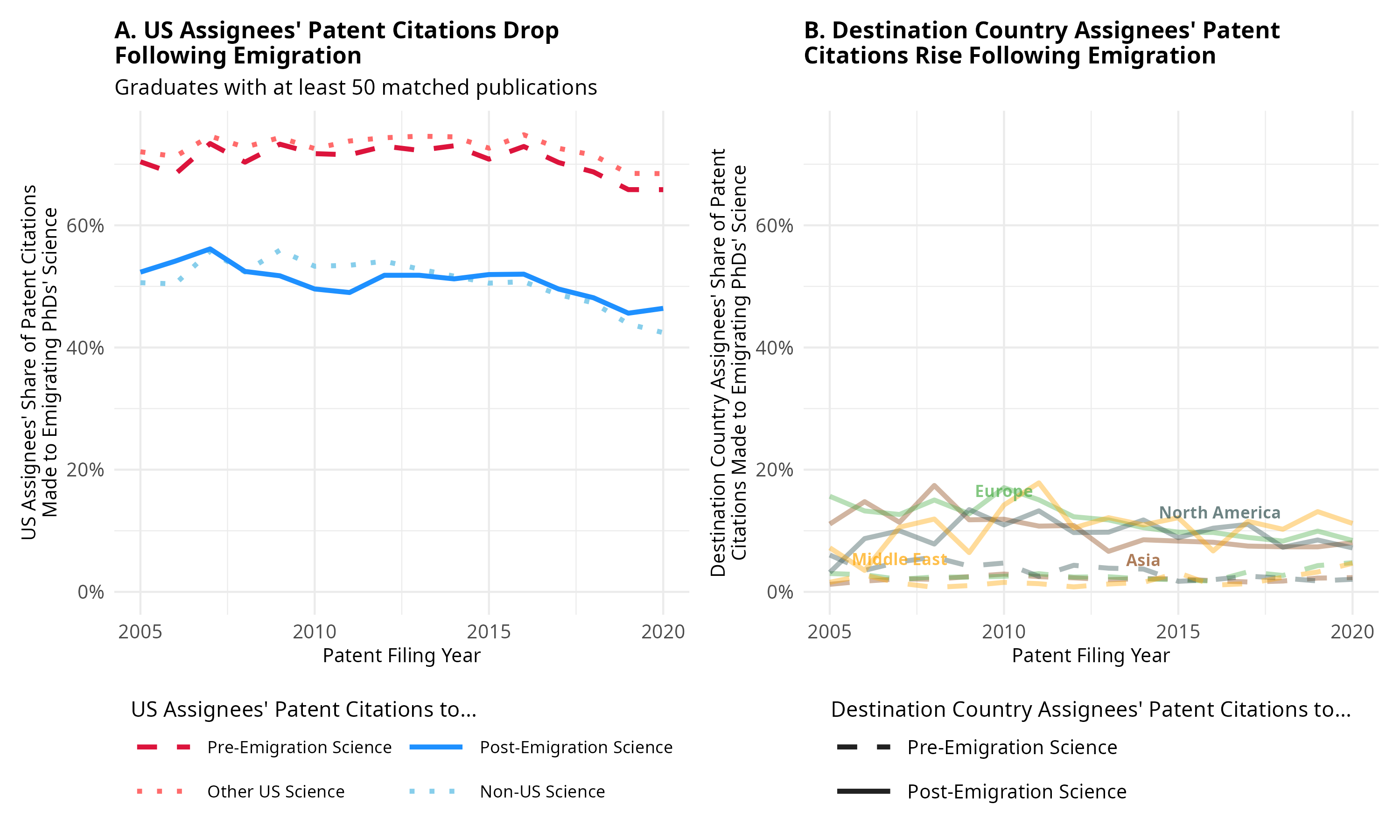} \\
\caption{\textbf{This figure presents a variant of Figure \ref{fig:figure2}, conditioning the sample to graduates with at least 50 linked publications.} Panel A evaluates US assignees' share of citations made each year in global patents to PhD graduates' science published before vs. after they leave the US (dashed red and solid blue lines, respectively). For comparison, the figure also provides US assignees' share of citations to non-leaver graduates' science and non-US science (dashed blue and red lines). Panel B shows a countervailing effect in destination countries.}
\label{fig:figure2_50pubs}
\end{figure}

\begin{figure}[H]
\centering
\includegraphics[width=0.8\linewidth]{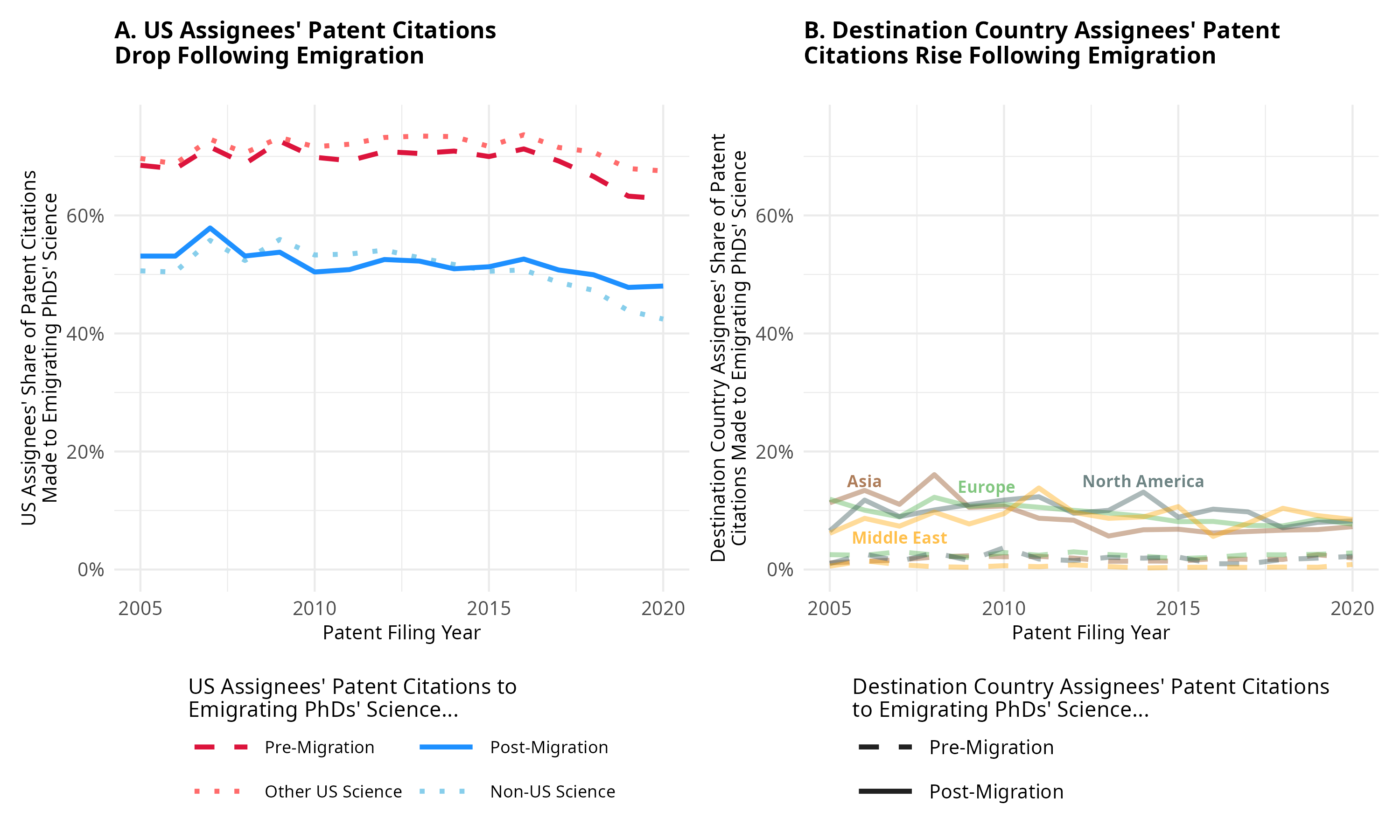} \\
\caption{\textbf{This figure presents a variant of Figure \ref{fig:figure2}, except rather than comparing patent citations to graduates pre vs. post-emigration science (as in Figure \ref{fig:figure2}), it compares patent citations made before vs. after the graduate emigrates.}} 
\label{fig:figure2_prepost}
\end{figure}

\begin{figure}[H]
\centering
\includegraphics[width=0.8\linewidth]{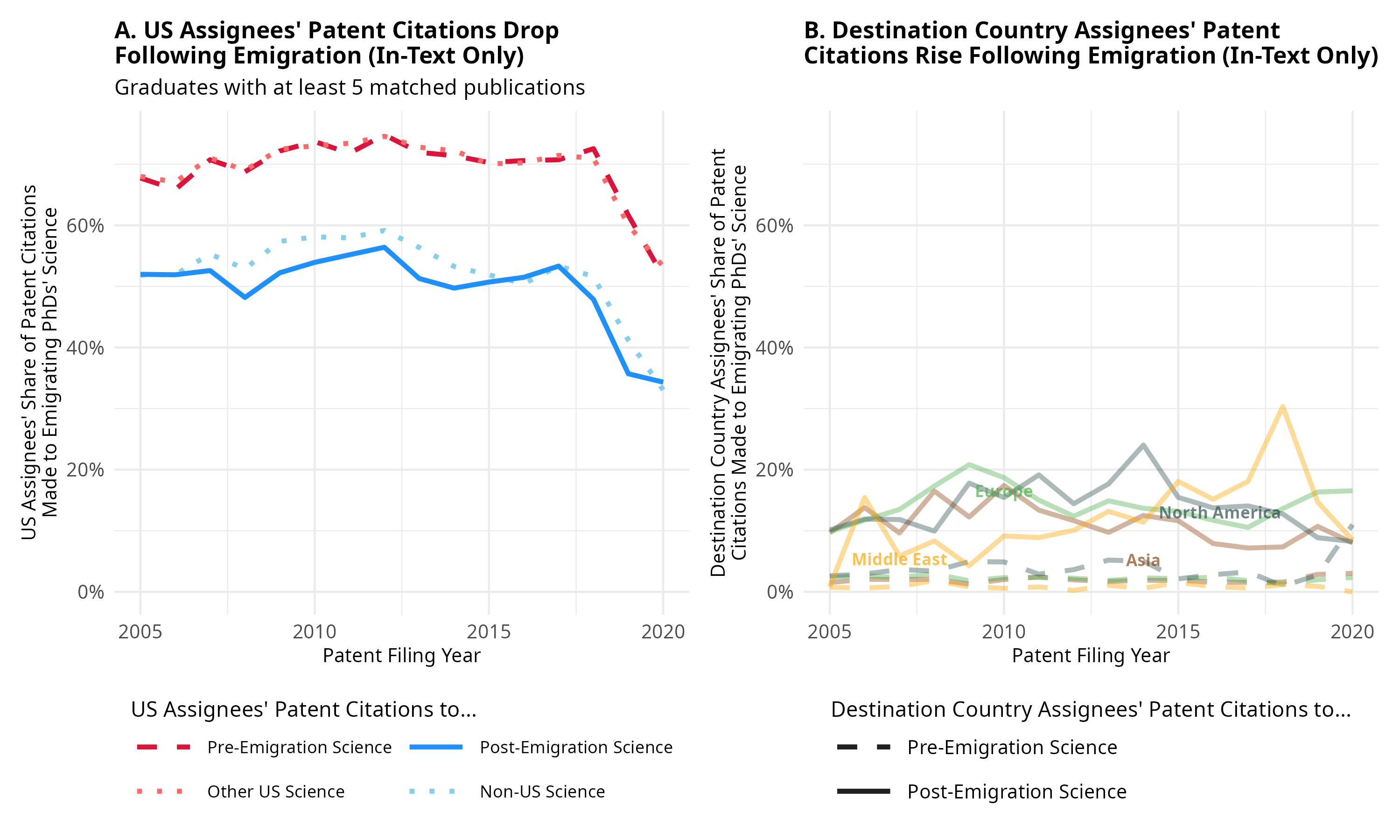} \\
\caption{\textbf{This figure presents a variant of Figure \ref{fig:figure2}, restricting patent citations to those made in the text of the patent specification (excluding ``front-page'' references).}} 
\label{fig:figure2_intext}
\end{figure}

Figures \ref{fig:figure2_10pubs} and \ref{fig:figure2_50pubs} similarly reproduce Figure \ref{fig:figure2} restricting to graduates with at least 10 and 50 linked publications, respectively, with similar findings. In Figure \ref{fig:figure2_prepost} we evaluate a variant on Figure \ref{fig:figure2}, where instead of comparing citations made to emigrants' pre- vs. post-emigration science (indexing on the date of publication), we compare citations made before vs. after emigration (indexing on the date of citation). The motivation for this test is that it is ex-ante ambiguous whether it is the location of the scientist at the time the work was produced versus their location at the time the work is referenced or used that matters more for its absorption. In the end, we find similar results either way. In Figure \ref{fig:figure2_intext}, we reproduce Figure \ref{fig:figure2} restricting patent citations to those made in the patent specification rather than in prior art references. Bryan et al. \citeApp{bryan2020text} argue that ``in-text'' citations may better measure an inventions' use of specific knowledge inputs than ``front-page'' (prior art) citations to non-patent literature (NPL), whose legal purpose is identifying prior art against which claims are evaluated, and may thus be a better indicator of intellectual proximity than of use per se. Our results in this figure remain similar.

\subsubsection*{Part 1: Extensions on Main Results}

\subsubsection*{Part 2: Robustness checks}

\pagebreak

\subsection*{Supplementary Tables}

\begin{table}[H]
\centering
\caption{\textbf{Stayers are more productive and higher impact than leavers.} Table estimates differences in between stayers and leavers in their (i) number of publications, (ii) average science citations per publication, (iii) average patent citations per publication, and (iv) patent-paper pairs. These features are measured using data only from each graduate's first five post-PhD years, and outcome is measured as an indicator for whether the graduate remained in the US for at least ten years. Sample consists of graduates between 2000 and 2014 to ensure at least 10 years of post-graduation data. The estimating equation is $Y_{i} = \beta \cdot 1(\textrm{Stayed at least 10 years}) + \alpha_{s} + \gamma_{f} + \delta_{t} + \varepsilon_{i}$, where $s$, $f$, and $t$ index school (PhD institution), field, and year, and the regression includes university, field, and graduation year fixed effects. *, **, *** represent significance at the 0.1, 0.05, and 0.01 levels, respectively. Robust standard errors provided in parentheses.}
\label{tab:stayers}
\vspace{.75ex}{
  \begin{tabular}{l*{6}{c}}
  \\ \toprule
                                  &\multicolumn{1}{c}{}&\multicolumn{1}{c}{Ln(Science}&\multicolumn{1}{c}{Ln(Patent}&\multicolumn{1}{c}{}&\multicolumn{2}{c}{Any...}     \\
                                &\multicolumn{1}{c}{Ln(Pubs)}&\multicolumn{1}{c}{cites/pub)}&\multicolumn{1}{c}{cites/pub)}&\multicolumn{1}{c}{Ln(PPP/pub)}&\multicolumn{1}{c}{Pat. cites}&\multicolumn{1}{c}{PPP}\\
\midrule
1(Stayer)                       &       0.025***&       0.037***&       0.104***&       0.166***&       0.033***&       0.020***\\
                                &     (0.004)   &     (0.005)   &     (0.009)   &     (0.014)   &     (0.002)   &     (0.002)   \\
\midrule
N                               &      213362   &      212450   &      136005   &       30953   &      216098   &      216098   \\
$R^2$                           &        0.08   &        0.21   &        0.15   &        0.06   &        0.15   &        0.05   \\
Y mean                          &       2.113   &       3.591   &      -0.139   &      -1.933   &       0.642   &       0.156   \\

  \\ \bottomrule
  \end{tabular}
}
\end{table}

\pagebreak


\clearpage
\bibliographyApp{bibliography}
\bibliographystyleApp{bib_econ}


\end{document}